\documentclass[]{jaa}

\usepackage{times} 
\usepackage{latexsym,amssymb,amsmath,amsfonts}
\usepackage{rotating} 
\usepackage{float}
\usepackage{graphicx}
\usepackage[round]{natbib}
\usepackage{longtable}
\usepackage{url}
\usepackage{dcolumn}
\usepackage{textcomp}
\usepackage{multicol}
\usepackage{bm}
\usepackage{pdflscape}
\usepackage[T1]{fontenc}
\usepackage{ae,aecompl}
\usepackage{appendix}
\usepackage[caption=false]{subfig}

\newcommand{\hi}{\mbox{H\,{\sc i}}}
\newcommand{\caii}{\mbox{Ca\,{\sc ii}}}
\newcommand{\nai}{\mbox{Na\,{\sc i}}}
\newcommand{\mgii}{\mbox{Mg\,{\sc ii}}}
\newcommand{\mgi}{\mbox{Mg\,{\sc i}}}
\newcommand{\feii}{\mbox{Fe\,{\sc ii}}}

\newcommand{\znii}{\mbox{Zn~{\sc ii}}}
\newcommand{\mnii}{\mbox{Mn~{\sc ii}}}
\newcommand{\crii}{\mbox{Cr~{\sc ii}}}

\def\h2{$\rm H_2$}
\def\Nh2{$N$(H${_2}$)}

\def\lymana{\ensuremath{{\rm Lyman}-\alpha}}
\def\kms{km\,s$^{-1}$}
\def\cms{cm$^{-2}$}
\def\cc{cm$^{-3}$}

\def\nhi{$N$($\hi$)}

\def\21{21-cm}
\def\ts{$T_{\rm s}$}

\def\fc{$C_{\rm f}$}

\def\taudv{$\int\tau dv$}
\def\ha{H\,$\alpha$}

\def\taudv{$\int\tau {\rm dv}$}
\def\c21{$C_{21}$}
\def\t0{$\tau_0$}
\def\wmg{$W_{\mgii}$}
\def\wfe{$W_{\feii}$}

\def\ebv{$E(B-V)$}

\begin{document}\sloppy

\title{Cold neutral hydrogen gas in galaxies}


\author{Rajeshwari Dutta\textsuperscript{1}}
\affilOne{\textsuperscript{1} European Southern Observatory, Karl-Schwarzschild-Str. 2, D-85748 Garching Near Munich, Germany\\}


\twocolumn[{

\maketitle

\corres{rdutta@eso.org}

\msinfo{}{}

\begin{abstract}
This review summarizes recent studies of the cold neutral hydrogen gas associated with galaxies probed via the \hi\ \21\ absorption line.
\hi\ \21\ absorption against background radio-loud quasars is a powerful tool to study the neutral gas distribution and kinematics
in foreground galaxies from kilo-parsec to parsec scales. At low redshifts ($z<0.4$), it has been used to characterize the distribution
of high column density neutral gas around galaxies and study the connection of this gas with the galaxy's optical properties. The neutral 
gas around galaxies has been found to be patchy in distribution, with variations in optical depth observed at both kilo-parsec and parsec 
scales. At high redshifts ($z>0.5$), \hi\ \21\ absorption has been used to study the neutral gas in metal or \lymana\ absorption-selected
galaxies. It has been found to be closely linked with the metal and dust content of the gas. Trends of various properties like incidence,
spin temperature and velocity width of \hi\ \21\ absorption with redshift have been studied, which imply evolution of cold gas properties 
in galaxies with cosmic time. Upcoming large blind surveys of \hi\ \21\ absorption with next generation radio telescopes are expected to 
determine accurately the redshift evolution of the number density of \hi\ \21\ absorbers per unit redshift and hence understand what drives 
the global star formation rate density evolution. 
\end{abstract}


}]


\doinum{12.3456/s78910-011-012-3}
\artcitid{\#\#\#\#}
\volnum{000}
\year{0000}
\pgrange{1--}
\setcounter{page}{1}
\lp{1}

\section{Introduction}
\label{sec:intro}
The interstellar medium (ISM) of a galaxy is a complex and dynamic system with constant interplay of matter, momentum and energy between the various ISM phases and the stars.
The phase structure of the ISM is hence a direct probe of various complex astrophysical processes that take place inside a galaxy. In addition, the ISM mediates between the 
stellar- and the galactic-scale processes. The accreting ionized gas from the intergalactic medium (IGM) passes though the different ISM phases before getting converted to 
stars. On the other hand, galactic winds, outflows and fountains populate the IGM with metals. Diffuse ionized hydrogen gas and metals have been observed to extend up to few 
hundreds of kilo-parsec around galaxies, constituting the circumgalactic medium \citep[CGM; ][]{tumlinson2017}. The large-scale multi-phase gas distribution in the CGM around 
galaxies, due to infalls, outflows and mergers, is again intricately linked with the ISM and the star formation in galaxies. Therefore, the ISM and CGM phases are expected to 
contain an imprint of the collective outcome of all the processes that shape the star formation history of the Universe \citep{hopkins2006}. 

In this review, we focus on the neutral gas phase in and around galaxies. In particular, we review studies of the cold neutral hydrogen gas as probed using the tool of \hi\ \21\ absorption. 
In Sections~\ref{sec:neutral} and \ref{sec:21cm}, we introduce the neutral gas phase and \hi\ \21\ absorption, respectively. We discuss results from studies of distribution of cold 
neutral gas around low-$z$ galaxies in Section~\ref{sec:galaxy}, dependence of cold neutral gas on metals and dust at high-$z$ in Section~\ref{sec:absorption}, and the redshift 
evolution of cold gas properties in Section~\ref{sec:evolution} We conclude by discussing future prospects in Section~\ref{sec:conclusion}

\section{The neutral gas phase}
\label{sec:neutral}
The neutral gas phase in and around galaxies act as the intermediary phase between the accreting ionized gas from the IGM and the molecular gas phase in the stellar disc that 
gets converted to stars. It thus is the reservoir from which molecules and stars form and consequently plays a crucial role in galaxy formation and evolution. The \hi\ disc 
around galaxies is the component that gets affected the most by tidal interactions and mergers, since it is more extended than the stellar disc, typically by more than a factor 
of two \citep{haynes1979,rosenberg2002,oosterloo2007,sancisi2008,chung2009,mihos2012}. Therefore, in the hierarchical structure formation models, galaxy formation is expected 
to leave its imprint on the atomic gas mass density and its evolution with redshift. 

The neutral gas exists in two dominant phases, the cold neutral medium (CNM; $T\sim100$~K, $n\sim30$\,\cc) and the warm neutral medium (WNM; $T\sim10^4$~K, $n\sim0.3$\,\cc). 
\citet{field1969} demonstrated that the CNM and the WNM can coexist in pressure equilibrium, such that the neutral gas can be considered as a two-phase medium. Observations 
of the Milky Way and nearby galaxies also provide evidence for a two-phase neutral medium \citep[e.g.][]{dickey1983,kulkarni1987,wolfire1995,mebold1997,braun1997,dickey2000}. 
\citet{wolfire1995} and \citet{wolfire2003} investigated the thermal balance of the WNM and CNM phases in the Galactic ISM. They found that their two-phase model, with thermal 
pressure in the range $\sim10^{3-4}$ K \cc, is in good agreement with observations of the Galactic ISM. In addition, the pressure equilibrium in their model depends on the gas 
phase abundance, dust abundance, metallicity, absorbing column density and radiation field. However, note that \citet{heiles2003} found that a large fraction, i.e. about 50\% 
of the WNM lies in the thermally unstable region of 500$-$5000\,K. The heating mechanisms in the neutral ISM are dust photoelectric heating, cosmic ray heating, X-ray heating 
and heating by C\,{\sc i} ionization, with dust photoelectric heating being the dominant heating mechanism. Radiative cooling by emission of the fine-structure lines, C\,{\sc ii}* 
$\lambda$158 $\mu$m and O\,{\sc i}* $\lambda$63 $\mu$m, dominates at $n\gtrsim1$ \cc\ ($T\lesssim$ 2000 K), while electron recombination onto positively charged grains and radiative 
losses from resonant transitions like Lyman-$\alpha$ dominate at $n\lesssim0.1$ \cc\ ($T\gtrsim$ 7000 K). 

\section{\hi\ \21\ absorption}
\label{sec:21cm}
The neutral gas in the ISM can be detected and studied via the \hi\ \21\ line, which originates in the hyperfine splitting of the $1s$ electronic ground state of hydrogen atom. 
Predicted by H.C. Van de Hulst in 1945 \citep{vandehulst1945}, \hi\ \21\ emission from our Galaxy was first detected in 1951 \citep{ewen1951,muller1951,pawsey1951}. Since then 
the \hi\ \21\ emission line has proved to be a powerful tool for studying the distribution, amount, kinematics, and temperature of the neutral gas in our Galaxy as well as other 
galaxies. In the local Universe ($z\lesssim$0.2), blind \hi\ \21\ emission-line surveys using single-dish telescopes have provided reliable measurements of the cosmological \hi\ 
mass density \citep{rosenberg2002,zwaan2005a,hoppmann2015,jones2018}, and spatially resolved \hi\ imaging using interferometers have traced the large scale dynamics of galaxies 
\citep{vanderhulst2001,zwaan2001,verheijen2007,deblok2008,walter2008,begum2008,catinella2015}. Rotation curves derived from \hi\ \21\ emission maps of galaxies have provided 
evidence for the existence of dark matter \citep{bosma1978,bosma1981a,bosma1981b,vanalbada1985,rubin1985,begeman1987,broeils1992}. However, the flux of the \hi\ \21\ emission 
signal is inversely proportional to the square of the distance to the unresolved emitting gas. Hence, sensitivities of present day radio telescopes make it difficult to directly 
map \hi\ emission from $z\gtrsim$0.2 galaxies. The highest redshift ($z$ = 0.376) \hi\ \21\ emission detection to date \citep{fernandez2016} has been possible due to very long 
integration (178 hours) using the Karl G. Jansky Very Large Array (VLA). Besides direct detection of \hi\ \21\ emission from galaxies, there have been several \hi\ \21\ emission 
line stacking experiments which have studied the average \hi\ properties of different samples of galaxies \citep{lah2007,lah2009,delhaize2013,rhee2013,rhee2018,kanekar2016}.
In addition to stacking, intensity mapping and cross-correlation of \hi\ \21\ emission with optical galaxy surveys \citep[e.g.][]{chang2010,masui2013} can complement individual 
\hi\ line surveys and probe the average \hi\ emission of galaxies upto higher redshifts ($z\sim$1).

Unlike \hi\ \21\ emission, the detectability of \hi\ \21\ absorption is not limited by distance of the absorbing gas and depends only on the strength of the background radio sources 
and \hi\ \21\ absorption cross-section projected on the sky by the galaxies. Therefore, \hi\ $21$-cm absorption line studies can complement the emission line surveys to trace the 
evolution of the atomic gas component in galaxies. The first Galactic detections of \hi\ \21\ absorption were reported by \citet{hagen1954a}, \citet{hagen1954b} and \citet{hagen1955}. 
\cite{clark1962} was the first to use extragalactic radio sources to probe Galactic \hi\ \21\ absorption, followed by \citet{shuter1969} and \citet{heiles1970}. The first detection 
of extragalactic \hi\ \21\ absorption was from the radio galaxy, Centaurus A \citep{roberts1970}. Thereafter, \citet{brown1973} detected \hi\ \21\ absorption at $z$ = 0.692 towards 
the background quasar 3C 286. Subsequently, \hi\ \21\ absorption has been used to probe the atomic gas in galaxies upto higher redshifts \citep[see][for a compilation]{kanekar2014a}, 
starting with detections of \hi\ \21\ absorption at $z>1$ by \citet{wolfe1979}, \citet{wolfe1981} and \citet{wolfe1985}. 

Further, \hi\ \21\ emission line observations of nearby dwarf and spiral galaxies indicate that properties of the CNM phase and $\sim$100\,pc- to 2\,kpc-scale structures detected in 
the \hi\ gas are closely linked with the in-situ star formation in galaxies \citep[e.g.][]{tamburro2009,bagetakos2011,ianjamasimanana2012}. However, identification of CNM gas through 
\hi\ \21\ emission is not straightforward as it depends on Gaussian decomposition of the emission line profiles. In the absence of absorption line measurements, it is not known whether 
the \hi\ \21\ emission line components exhibiting smaller line widths (and hence assumed to correspond to CNM) are truly cold. Therefore, the contributions due to turbulent motions and 
the processes driving the observed properties of \hi\ gas are poorly constrained. Moreover, \hi\ \21\ emission studies usually do not have sufficient spatial resolution to detect 
parsec-scale structures. On the other hand, \hi\ \21\ absorption is an excellent tracer of the CNM phase \citep{kulkarni1988}, and can be used to study parsec-scale structures in the 
\hi\ gas using sub-arcsecond-scale spectroscopy \citep[e.g.][]{srianand2013}. 

The \hi\ \21\ optical depth integrated over velocity ($\int{\tau {\rm dv}}$ in \kms) is related to the column density of neutral hydrogen, \nhi\ (\cms), spin temperature \ts\ (K) 
of the gas, and the fraction \fc\ of the background radio source covered by the absorbing gas, as \citep[e.g.][]{rohlfs2000}: 
\begin{equation}
 N(\hi) = 1.823 \times 10^{18}~ \frac{T_{\rm s}}{C_{\rm f}} \int{\tau {\rm dv}},
\label{chapter1_eqn_nhi}
\end{equation}
where the \hi\ \21\ line is assumed to be optically thin.  
Note that if there are a number of \hi\ gas clouds at different phases along the line-of-sight giving rise to the absorption line, then the effective \ts\ measured would be the 
column density-weighted harmonic mean of the \ts\ of the different clouds.
The inverse dependence of the \hi\ \21\ optical depth on \ts, coupled with its very low transition probability and its resonance frequency falling in the radio 
wavelengths, make the \hi\ \21\ absorption line a good tracer of high column density cold \hi\ gas without being affected by dust and luminosity biases.
Hence, the \hi\ \21\ absorption line can be used to investigate: 
\begin{enumerate}
 \item the thermal state of the \hi\ gas, since the \hi\ \21\ optical depth depends inversely on \ts, which is known to follow the gas kinetic temperature in the CNM 
       \citep{field1959,bahcall1969,mckee1977,wolfire1995,roy2006}, and can be coupled to the gas kinetic temperature via the \lymana\ radiation field in the WNM 
       in case of a two-phase medium \citep{liszt2001};
 \item the kinetic temperature of the gas can also be constrained using the thermal widths of the individual \hi\ \21\ absorption components, subject to the Gaussian modeling 
       \citep[e.g.][]{lane2000,kanekar2001};
 \item the parsec-scale structure in the absorbing gas via sub-arcsecond-scale spectroscopy \citep[e.g.][]{srianand2013,biggs2016,gupta2018a};
 \item the magnetic field in the CNM using Zeeman splitting \citep{heiles2004}; 
 \item the filling factor of cold gas in the ISM/CGM of galaxies; 
 \item the underlying potential using the overall velocity width of the \hi\ \21\ absorption lines;
 \item the temporal and spatial variation in fundamental constants of physics like electromagnetic fine structure constant, the proton-to-electron mass ratio and the proton g-factor 
       \citep{wolfe1976,carilli2000,chengalur2003,kanekar2010,kanekar2012,rahmani2012}.
\end{enumerate}

Next, we discuss some of the limitations of the \hi\ \21\ absorption line technique. The main limitation when estimating \nhi\ from \hi\ \21\ absorption is that the two parameters $-$ 
\ts\ and \fc\ in Eqn.~\ref{chapter1_eqn_nhi}, cannot usually be well-constrained. Comparison of both \hi\ \21\ emission and absorption spectra can be used to estimate \ts. Such studies 
are possible in our Galaxy and nearby galaxies \citep[e.g.][]{wakker2011,keeney2011,dutta2016}. Though it should be noted that the angular scales probed by \hi\ \21\ emission are usually 
larger than that by \hi\ \21\ absorption. Therefore, \nhi\ obtained from \hi\ emission maps represents average \hi\ surface density, while the cold gas traced by \hi\ \21\ absorption 
could be much clumpier, implying that there will be uncertainties in comparing \hi\ emission to absorption. However, such comparisons of \hi\ emission and absorption are not feasible at 
high redshifts as explained above. Hence, for identifying and studying the ISM of high redshift galaxies, one usually has to rely on absorption lines detected towards bright background 
sources like quasars and gamma ray bursts. In such cases, \ts\ can be estimated if there is independent measurement of \nhi\ from \lymana\ absorption in optical or ultraviolet (UV) spectra. 
However, this technique suffers from uncertainties regarding how much of the \hi\ gas traced by \lymana\ absorption is associated with the \21\ absorbing gas, and whether the radio and 
optical lines-of-sight are aligned. The second parameter, \fc, is usually taken to be unity for lack of sufficient information about the extent of the absorber and the background radio 
source. This can be estimated by comparing the flux density of the radio source at high spatial resolution (parsec-scales), from Very Long Baseline Interferometry (VLBI) observations at 
the redshifted \hi\ \21\ line frequency, with the total flux density at the lower spatial resolution (usually kilo-parsec-scales) at which absorption is observed 
\citep[e.g.][]{kanekar2009b,gupta2012}. However, VLBI observations of the radio sources may not always be available, especially at lower frequencies as required for the high redshift absorbers.

\begin{figure}
\includegraphics[width=0.45\textwidth]{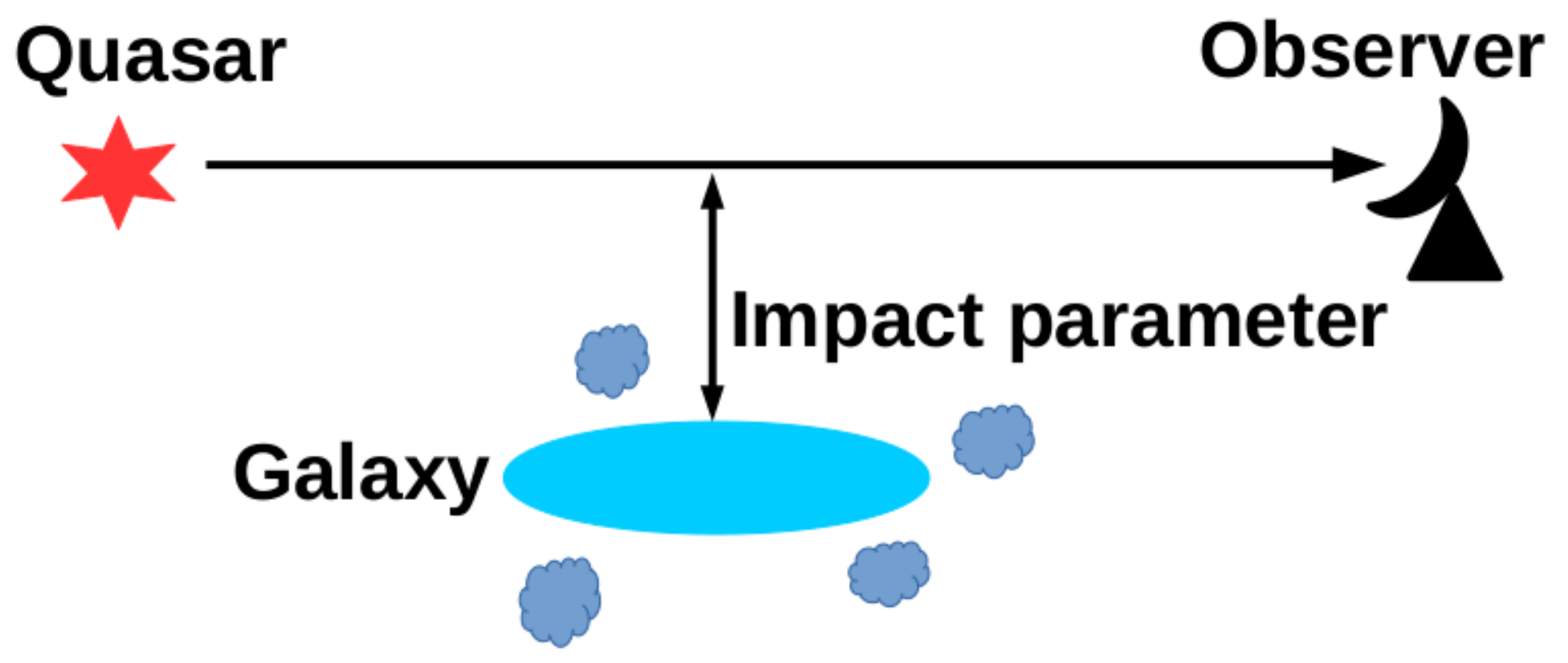}
\caption{Schematic diagram to illustrate how spectra of background quasars are used to probe the gas around foreground galaxies. The projected separation between the quasar and galaxy 
is termed as the impact parameter.}
\label{fig:quasar_schematic}
\end{figure}

\hi\ \21\ absorption can be classified into two categories: (a) intrinsic and (b) intervening. The former probes neutral gas associated with the Active Galactic Nucleus (AGN) 
and its environment, and can be used to constrain models of formation and evolution of AGN, as well as the feedback they provide to their host galaxies \citep[see for a review][]{morganti2018}. 
The latter probes the neutral gas in foreground galaxies and can be used to study how galaxies form and evolve with time. Here we concentrate on the latter category. 
See Fig.~\ref{fig:quasar_schematic} for an illustration of quasar absorption line technique to study the gas associated with intervening galaxies.

\section{Distribution of cold neutral gas around low-$z$ galaxies}
\label{sec:galaxy}
Absorption lines seen in the spectra of background quasars whose sightline happen to pass through the discs or halos of foreground galaxies (we refer to such fortuitous associations 
as quasar-galaxy-pairs or QGPs from hereon), allow us to probe the physical, chemical and ionization state of gas in different environments such as the stellar discs, extended \hi\ 
discs, high velocity clouds, outflows, accreting streams and tidal structures. The main drawback of quasar absorption line spectroscopy is that it probes the gas only along the pencil 
beam sightline. This issue can be addressed by compiling a large homogeneous and statistically significant sample of absorbers. The relationship between absorption strength and impact 
parameter (projected separation between quasar sightline and galaxy centre) obtained from a large number of quasar sightlines passing near foreground galaxies can then be used to 
statistically determine the gas distribution in and around galaxies. Considerable progress has been made in mapping the distribution of gas in the CGM or galaxy halos using absorption
from \lymana\ \citep{chen2001,prochaska2011,tumlinson2013,stocke2013,borthakur2015}, \mgii\ \citep{chen2010a,kacprzak2012,churchill2013,nielsen2013,bordoloi2014a}, and other metals 
like C\,{\sc ii}, C\,{\sc iii}, C\,{\sc iv}, Si\,{\sc ii}, Si\,{\sc iii}, Si\,{\sc iv}, O\,{\sc vi} \citep{tumlinson2011,werk2014,bordoloi2014b,liang2014}. 
Such studies have shown that the CGM can extend upto the virial radius ($\sim$100s of kpc); is typically cool ($\lesssim$10$^5$ K, i.e. below the virial temperature); is likely to be 
bound to the dark matter halo of the galaxy (within $\sim \pm$200\,\kms\ of the galaxy); is multiphase in nature; and becomes progressively more ionized with increasing distance from 
the galaxy centre. In addition to the above, the \lymana\ absorbing gas is found to be ubiquitous around both star-forming and passive galaxies, while the metals are found to be more 
patchy in distribution.

Here we are interested in probing the high column density \hi\ gas around galaxies. Such gas is usually traced by damped \lymana\ absorbers (DLAs; \nhi $\ge$ 2$\times$10$^{20}$\,cm$^{-2}$)
and sub-DLAs (\nhi $\sim$ 10$^{19}$ $-$ 2$\times$10$^{20}$\,cm$^{-2}$) \citep[see][]{wolfe2005,peroux2005}. Thanks to large spectroscopic surveys like the Sloan Digital Sky Survey 
\citep[SDSS;][]{york2000}, thousands of DLAs are known at $z>$ 1.8, and they are found to trace the bulk ($\sim$80\%) of the \hi\ gas at 2 $\le z\le$ 4 
\citep{prochaska2005,noterdaeme2009b,noterdaeme2012b}. However, atmospheric cutoff of light below 3000 \AA\ restricts ground-based observations of DLAs at $z<$ 1.5. UV spectroscopic 
observations with the Hubble Space Telescope (HST) have identified $\sim$70 low-$z$ ($z<$1.65) DLAs and sub-DLAs till date 
\citep{rao2006,rao2011,meiring2011,battisti2012,turnshek2015,neeleman2016,rao2017}.
Using ground-based imaging studies of $z<$ 1 DLAs, \citet{rao2011} have found an anti-correlation (Spearman rank correlation coefficient, $r_s$ = $-$0.34 at 3$\sigma$ level of significance) 
between \nhi\ and impact parameter ($b$), with median $b$ = 17 kpc. They also do not find any correlation between galaxy luminosity and \nhi\ and no significant evidence that galaxies at 
larger $b$ are more luminous. Similar trend of \nhi\ decreasing with $b$ is found by \citet{rahmani2016} using X-Shooter observations of $z\sim$0.6 DLA host galaxies and by \citet{peroux2012} 
using integral field unit observations of \ha\ emission from DLAs at 1 $<z<$ 2. At $z>$ 2, \citet{krogager2012} have found an anti-correlation ($r_s$ = $-$0.6) between \nhi\ and $b$, as well as 
correlation between metallicity and $b$, consistent with simulations that contain feedback mechanisms to control the star formation \citep{fynbo2008,pontzen2008}. 

At $z=$ 0, \citet{zwaan2005}, using \hi\ \21\ emission maps of local galaxies, have found that \nhi\ decreases with galactocentric radius. Further, they have calculated the two-dimensional
probability function of \hi\ column density and impact parameter, and found that the distribution of $b$ and luminosities of $z<$ 1 DLA host galaxies can be explained on the basis of 
the local galaxy population. Based on their conditional probability distribution of \nhi\ and $b$, the expected median $b$ for systems with log~\nhi\ $>$ 20.3 (\cms) is 8 kpc. However, 
\citet{rao2011} found that a significantly higher fraction of low-redshift DLA host galaxies are at larger $b$ values, with luminosities less than the characteristic luminosity, which 
may hint at an evolution in the \hi\ sizes of galaxies with redshift. Similar studies of \hi\ \21\ emission in nearby dwarf galaxies have found that their \hi\ column density distribution 
function falls off significantly faster at high \nhi\ compared to that in DLAs and the local luminous galaxy population \citep{begum2008,patra2013}. Hence, the extent of high \nhi\ gas 
around them is likely to be much smaller and in order to detect gas with high \nhi\ from dwarf galaxies one must probe them at very small impact parameters. 

While \lymana\ absorption and \hi\ \21\ emission trace the neutral gas around galaxies, \hi\ \21\ absorption towards radio-loud quasars is an excellent tracer of the CNM phase in galaxies.
A detailed study of cold \hi\ gas in the outer disks and halos of galaxies is required to understand where and how the infalling/circumgalactic gas condenses into the ISM. There are usually 
two approaches to map the distribution of gas around galaxies using quasar absorption lines: (i) galaxy-blind or absorption-selected approach, where one searches for intervening absorption 
towards a background quasar and then tries to identify the associated host galaxy, and (ii) absorption-blind or galaxy-selected approach, where one selects a galaxy in close proximity to a 
background quasar (i.e. a QGP) without prior knowledge of any absorption along the quasar sightline, and then proceeds to search for absorption at the redshift of the galaxy towards the quasar. 
The advantage of the second approach over the first is that it is not biased against dusty sightlines (which are more likely to be conducive to the presence of cold gas), and hence it is expected
to trace gas associated with the general galaxy population without any biases.

Following the second approach, \hi\ \21\ absorption searches from low-$z$ ($z<$0.4) QGPs have revealed a weak anti-correlation between the \hi\ \21\ optical depth and impact parameter 
\citep{carilli1992a,gupta2010,borthakur2011,borthakur2016,zwaan2015,reeves2016}. However, the number of low-$z$ QGPs in these studies was too small to characterize the distribution of 
cold \hi\ gas around galaxies. With a view to map the distribution of high column density (\nhi\ $\ge$ 10$^{19}$ \cms) cold ($T\sim$ few 100 K) \hi\ gas around low-$z$ galaxies, 
\citet{dutta2017a} have carried out a systematic survey of \hi\ \21\ absorption in a homogeneous sample of 55 $z<0.4$ QGPs (40 of which are in the statistical sample) towards radio 
sources at $b$ $\sim$0$-$35 kpc. The main results from this study are highlighted below.

\begin{figure*}[t]
\centering \includegraphics[width=0.5\textwidth, angle=90]{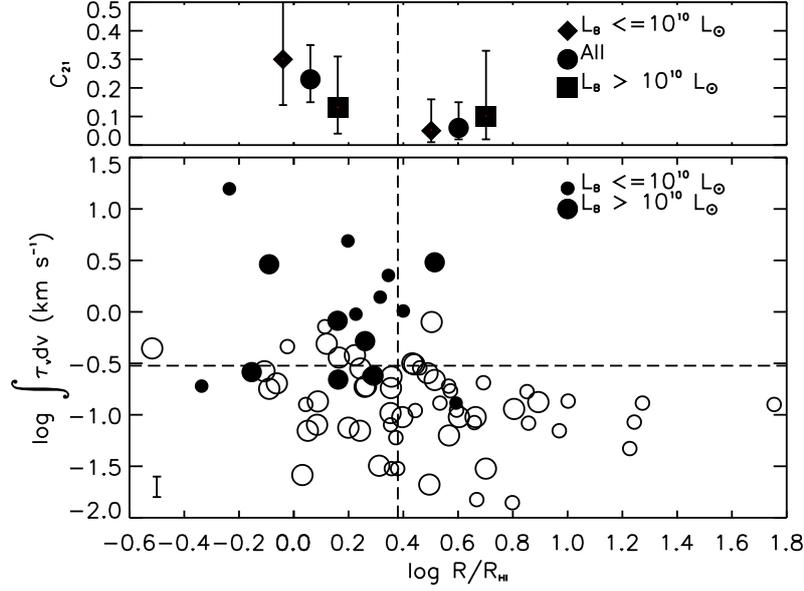}
\caption{{\it Bottom:} Integrated \hi\ \21\ optical depth as a function of radial distance from galaxy's centre, scaled with the effective \hi\ radius \citep[see for details][]{dutta2017a}. 
The solid and open circles represent measurements and $3\sigma$ upper limits (for typical velocity width of 10\,\kms) of the \hi\ \21\ detections and non-detections, respectively. 
The small circles are for galaxies with luminosity, $L_B \le$ 10$^{10}$ $L_\odot$, while the large circles are for galaxies with $L_B >$ 10$^{10}$ $L_\odot$. The typical error in the 
optical depth measurements is shown at the bottom left of the plot. The horizontal dotted line marks the optical depth sensitivity used for the covering factor estimate, i.e. \taudv\ = 0.3\,\kms, 
and the vertical dotted line marks the median scaled radial distance.
{\it Top:} The covering factor or detection rate of \hi\ \21\ absorbers is shown in two different radial distance bins demarcated at the median value. The circles, diamonds and 
squares are for all the galaxies, galaxies with $L_B \le$ 10$^{10}$ $L_\odot$ and galaxies with $L_B >$ 10$^{10}$ $L_\odot$, respectively. {\it Both the optical 
depth and incidence of \hi\ \21\ absorption are weakly anti-correlated with distance from galaxy's centre.}}
\label{fig:radial_profile}
\end{figure*}

\begin{figure*}[t]
\centering \includegraphics[width=1.0\textwidth]{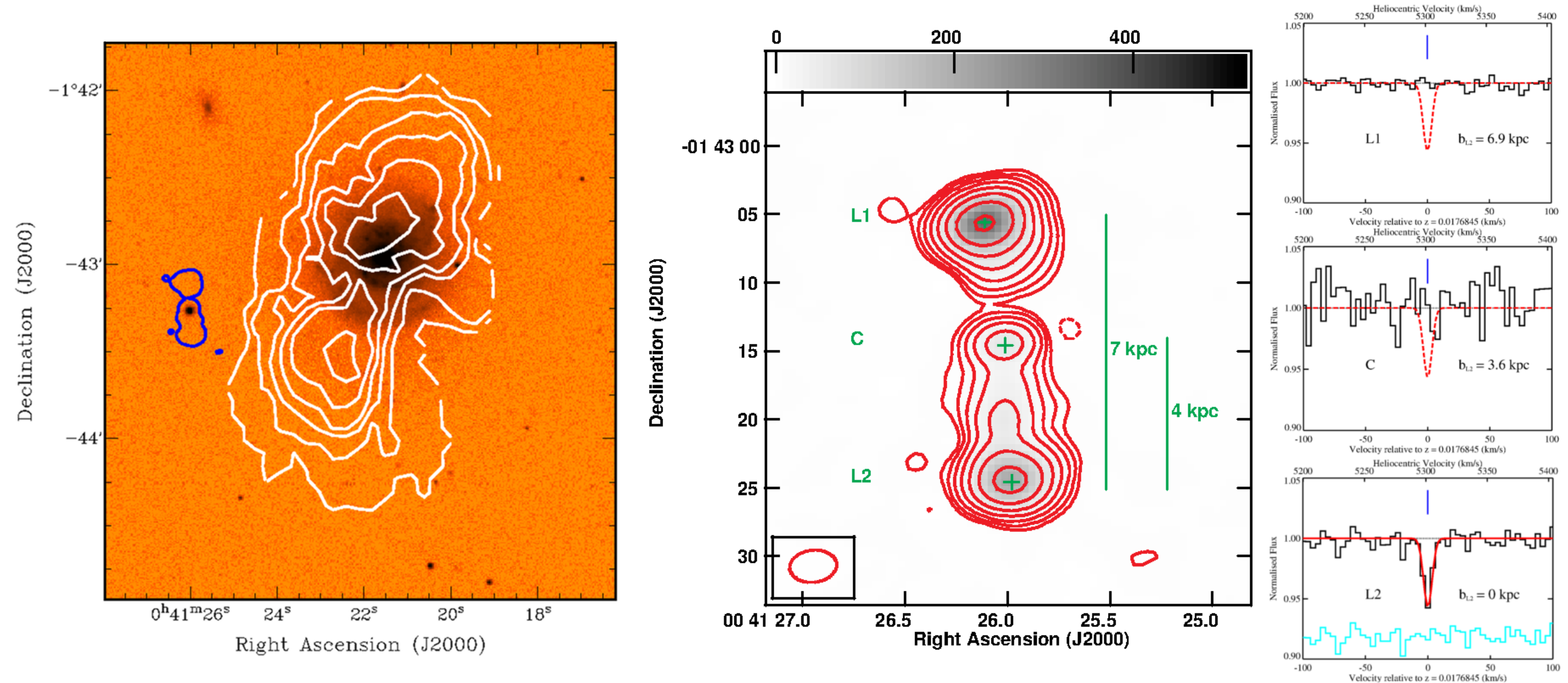}
\caption{{\it Left:} SDSS $r$-band optical image of a QGP, overlaid with the \hi\ emission contours of the foreground ($z$ = 0.02) galaxy in white, and the outermost contour of the 
1.4~GHz continuum of the background radio source in blue \citep[see for details][]{dutta2016}.
{\it Centre:} GMRT 1.4~GHz continuum contours of the radio source that is extended over 7~kpc at the redshift of the foreground galaxy. The contour levels are plotted as 2.5 $\times$ 
($-$1,1,2,4,8,...)\,mJy~beam$^{-1}$, where solid (dashed) lines correspond to positive (negative) values. At the bottom left corner of the image the restoring beam is shown as an ellipse. 
The continuum peaks of the core, the northern lobe and the southern lobe are identified as C, L1 and L2 respectively. The vertical lines indicate the projected separation between these 
components at the redshift of galaxy.
{\it Right:} GMRT \hi\ \21\ absorption spectra towards the different continuum components, as marked in the left panel. The best-fitting single Gaussian profile to the \hi\ \21 absorption 
towards L2 is overplotted in solid red line, and the residuals from the fit are plotted below in cyan. This fit is also overplotted on the spectra towards L1 and C. The vertical tick marks 
the position of the peak optical depth detected towards L2. {\it The \hi\ \21\ optical depth varies by a factor of $\ge$7 over 7~kpc at similar impact parameter of 25~kpc from the galaxy.}
}
\label{fig:mosaic}
\end{figure*}

\begin{enumerate}
 \item {\it Radial profile of cold \hi\ gas around low-$z$ galaxies:}
       The strength and covering factor of \hi\ \21\ absorption decreases, albeit slowly, with increasing impact parameter, radial distance along the galaxy's 
       major axis and distances scaled with the effective \hi\ radius (see Fig.~\ref{fig:radial_profile}). There is a weak anti-correlation (rank correlation 
       coefficient = $-$0.20 at 2.42$\sigma$ level) between \taudv\ and $b$. The covering factor of \hi\ \21\ absorbers decreases from 0.24$^{+0.12}_{-0.08}$ 
       at $b\le$ 15 kpc to 0.06$^{+0.09}_{-0.04}$ at $b=$ 15$-$35 kpc. The 3$\sigma$ \taudv\ sensitivity for this estimate is 0.3\,\kms, which corresponds 
       to \nhi\ = $5\times10^{19}$\,\cms, for a \ts\ of 100\,K typical of CNM and \fc\ of unity.
 \item {\it Azimuthal profile of cold \hi\ gas around low-$z$ galaxies:}
       There is tentative evidence that the distribution of the \hi\ \21\ absorbers is likely to be co-planar with that of the \hi\ disk. The absorption strength
       and covering factor are higher when the radio sightline passes near the galaxy's major axis. Further, the covering factor is maximum for sightlines that 
       pass near the major axis of edge-on galaxies.
 \item {\it Dependence of cold \hi\ gas distribution on host galaxy properties:} 
       The strength and covering factor of \hi\ \21\ absorbers is not found to depend significantly on the host galaxy properties, i.e. luminosity, stellar mass, 
       colour, surface star formation rate density and redshift. Hence, it is surmised that the distribution of \hi\ \21\ absorbers is more sensitive to geometrical 
       parameters than physical parameters related to the star formation in galaxies.
 \item {\it Nature of cold \hi\ gas around low-$z$ galaxies:} 
       No correlation is found between \hi\ \21\ optical depth and equivalent widths of \caii\ and \nai\ absorption lines detected in the optical spectra of the quasars. 
       The observed equivalent ratios of \caii\ and \nai\ suggest that most of the \hi\ \21\ absorbers observed around low-$z$ galaxies are not tracing the dusty star-forming 
       disks, but rather the diffuse extended \hi\ disks. Further, the observations suggest that cold gas clouds in the extended disks/halos of galaxies have small sizes 
       (parsec to sub-parsec scale) and are patchy in distribution (see Fig.~\ref{fig:mosaic}). Indeed, there have been observations of structures in the \hi\ gas around 
       galaxies from parsec-scales \citep{srianand2013,dutta2015} to kilo-parsec-scales \citep{dutta2016}. This is further supported by four times higher incidence of 
       \hi\ \21\ absorption around $z<1$ DLA host galaxies (i.e. absorption-selected sample) compared to the galaxy-selected sample of QGPs. From the fact that $\sim$60\% of 
       $z<1$ DLAs have cold gas that can produce detectable \hi\ \21\ absorption, we infer that the \hi\ gas distribution around low-$z$ galaxies that contribute to the DLA 
       population is patchy, with a covering factor of $\sim$30\% within $\sim$30 kpc.
\end{enumerate}


\section{Dependence of cold neutral gas on metals and dust}
\label{sec:absorption}
Currently it is not possible to extend the study of low-$z$ QGPs, as described in Section~\ref{sec:galaxy}, to higher redshifts (i.e. $z>0.5$) due to the difficulty of constructing 
large spectroscopic samples of galaxies at high redshifts. Hence, the usual practice to study cold gas at high redshifts has been to search for \hi\ \21\ absorption towards radio-loud
quasars that show strong metal or \lymana\ absorption in their optical and UV spectra, i.e. absorption-selected approach. As mentioned in Section~\ref{sec:galaxy}, the \lymana\ line
cannot observed from ground at $z<1.5$. The \mgii\ doublet lines, $\lambda\lambda$ 2797, 2803, offer the best way to probe high \nhi\ systems in the absence of direct observations of 
\lymana. \mgii\ absorption detected towards background quasars have proved to be excellent tools to probe the gaseous halos of $z\lesssim$2 galaxies  
\citep{lanzetta1987,sargent1988,bergeron1991,steidel1992,steidel1995,nestor2005,prochter2006,quider2011,zhu2013}. At 0.5$\le z\le$3.5, systematic searches of \hi\ \21\ absorption
in samples of \mgii\ systems and DLAs towards radio-loud quasars have estimated the detection rate or the CNM filling factor as 10$-$20\% 
\citep{briggs1983,kanekar2003a,curran2005,gupta2009,kanekar2009a,curran2010a,srianand2012,gupta2012,kanekar2013,kanekar2014a}. 
Spin temperature (\ts) measurements derived using \hi\ \21\ optical depth and \nhi\ measured from DLAs, suggest that most of the gas along these sightlines trace the diffuse WNM phase, 
and only a small fraction of the total \nhi\ is associated with the CNM phase \citep{srianand2012,kanekar2014a}. This is supported by observations of typically low ($\sim10-20$\%) 
molecular fractions of \h2\ in $z>1.8$ DLAs \citep{noterdaeme2008a}, as well as physical conditions inferred in DLAs using C\,{\sc ii}* and Si\,{\sc ii}* fine structure lines \citep{neeleman2015}.
In addition, there are indications for an anti-correlation between \ts\ and the gas phase metallicity \citep[see][and references therein]{kanekar2014a}.

Strong \mgii\ absorbers (rest equivalent width of \mgii\ $\lambda 2796$, \wmg\ $\ge$1\,\AA) at $z<$1.65 have been shown to trace gas with high neutral hydrogen column densities, like 
DLAs and sub-DLAs \citep{rao2006}. However, strong \mgii\ systems sample a wide range of galaxy impact parameters, over $\sim$10$-$200 kpc \citep{nielsen2013}. Hence, such absorbers 
can trace gas in a wide variety of environments like star-forming discs, CGM, galactic winds and outflows. To study the cold dense gas around galaxies, other parameters like equivalent 
width ratios of metal lines are required to select sightlines that probe low impact parameters. \citet{rao2006} have demonstrated that equivalent width ratios of \mgii, \mgi\ and \feii\ 
absorption can be used to pre-select DLAs more successfully than by just using \wmg. Further insights into the origin and physical conditions prevailing in the strong \mgii\ systems and 
DLAs can be obtained by studying their associated \hi\ \21\ absorption. \citet{gupta2009,gupta2012} have shown that the \hi\ \21\ detection rate in strong \mgii\ systems can be enhanced 
with appropriate equivalent width ratio cuts of \mgii, \feii\ and \mgi. Recently, \citet{dutta2017b} have proposed an efficient \feii\ absorption-based selection technique to detect high 
\nhi\ cold gas at high-$z$. The detection rate of \hi\ \21\ absorption increases with the absorption strength of \feii, and additional constraints on \feii\ (rest equivalent width of 
\feii\ $\lambda 2600$, \wfe\ $\ge$1\,\AA) gives a higher (by a factor of $\sim4$) detection rate of \hi\ \21\ absorption compared to a pure \mgii-based selection. 

\begin{figure*}[t]
\centering \includegraphics[width=1.0\textwidth]{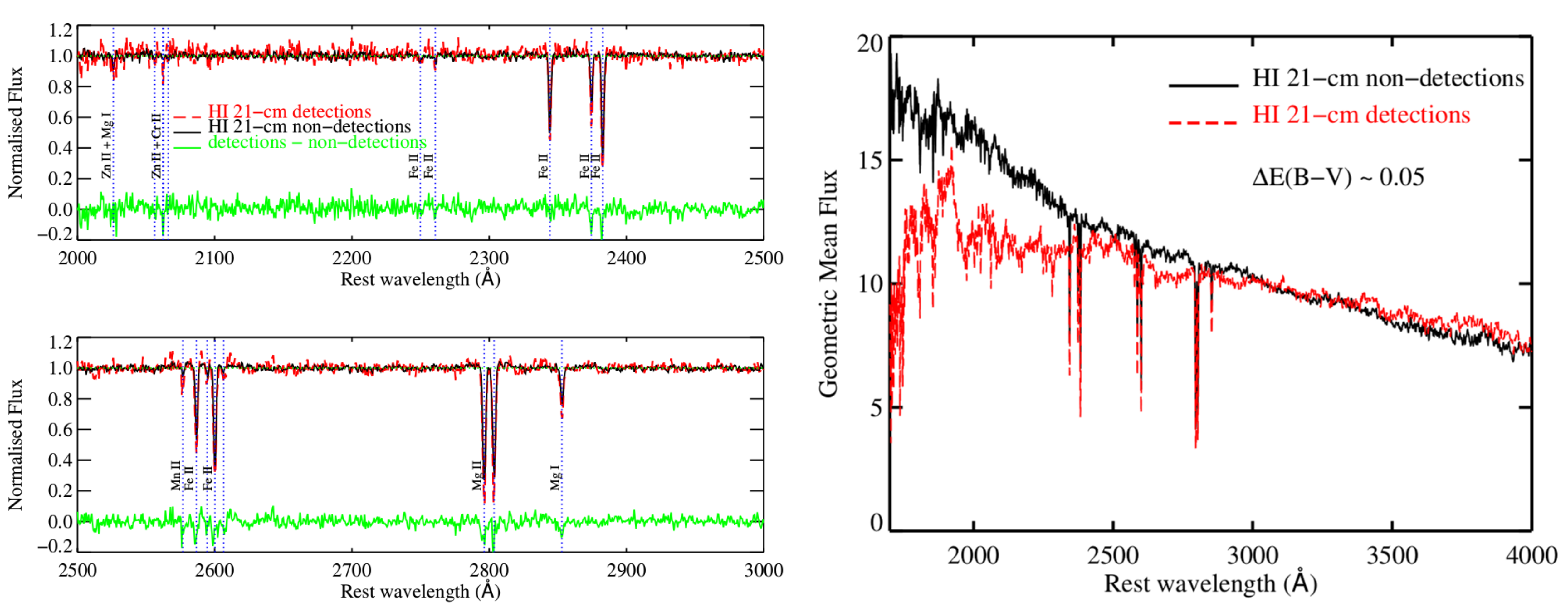}
\caption{{\it Left:} The median stacked spectrum of SDSS quasars at $0.5<z<1.5$ which are (not) detected in \hi\ \21\ absorption is shown as the red dashed (black solid) line
\citep[see for details][]{dutta2017b}. The difference of the stacked spectrum of \hi\ \21\ non-detections from that of the detections is shown as the green line at an 
arbitrary offset in the $y$-axis for clarity. The various rest wavelengths of transitions of \mgii, \mgi, \crii, \mnii, \feii\ and \znii\ are marked by vertical dotted lines. 
{\it The systems which show \hi\ \21\ absorption, also show systematically stronger metal absorption (i.e. larger equivalent widths of the metal lines by $\sim3-4\sigma$) than 
those which do not.} {\it Right:} The geometric mean stacked spectrum of SDSS quasars at $0.5<z<1.5$ which are (not) detected in \hi\ \21\ absorption is shown as the red dashed 
(black solid) line \citep[see for details][]{dutta2017b}. The differential reddening, $\Delta$\ebv, is 0.05. {\it \hi\ \21\ absorption on an average causes more reddening 
in the quasar spectra, indicating the presence of more dust.}}
\label{fig:sdss_stack}
\end{figure*}

Further, the properties of cold gas detected through \hi\ \21\ absorption are found to be closely linked with the metal and dust content of the gas. \citet{dutta2017b} have found that 
\hi\ \21\ absorption arises on an average in systems with stronger metal absorption. Stacking of SDSS optical spectra of background quasars shows that the average equivalent widths of 
various metal lines among \hi\ \21\ absorbers are higher by a factor of $\sim3-4$ than that found in non-absorbers (Fig.~\ref{fig:sdss_stack}, left panel). In addition, quasars with 
\hi\ \21\ absorption detected towards them are found to be systematically more reddened than those without absorption (Fig.~\ref{fig:sdss_stack}, right panel), and there is a tendency 
for the detection rate of \hi\ \21\ absorbers to be higher towards more reddened quasars. Further, \hi\ \21\ absorption searches towards radio-selected red quasars have usually 
resulted in a higher detection rate than that obtained for optically-selected DLAs \citep[e.g.][]{carilli1998,ishwara2003}. There have also been detections of \hi\ \21\ and molecular 
line absorption towards gravitationally-lensed systems that have high visual extinction \citep{carilli1992b,chengalur1999,kanekar2003b}. Hence, all the above imply that \hi\ \21\ 
absorption is more likely to arise in metal-rich dusty cold gas. 

\section{Redshift evolution of cold neutral gas}
\label{sec:evolution}
The global star formation rate density (SFRD) of the Universe peaks at $z\sim2$, followed by a decline towards $z=0$ \citep{madau2014}. The redshift evolution of the SFRD is expected to be 
imprinted in the ISM/CGM of galaxies, because the physical conditions and the volume filling factors of different gas phases depend on various feedback mechanisms associated with the in-situ 
star-formation. Therefore, mapping the redshift evolution of different gas phases will provide deeper understanding of the physical processes that drive the global star formation in the 
Universe. Even more relevant for understanding the SFRD evolution is how the fraction of cold gas (which acts as the reservoir for star-formation) is evolving with redshift, which is not 
well-constrained from observations. The cosmic mass density of \hi, $\Omega_{\hi}$, shows a mild (factor of $\sim2-3$) decrease from $2<z<4$ to $z\sim0.2$ \citep{rhee2018}. This is modest 
compared to an order of magnitude decrease in the SFRD over the same redshift range. This implies that the processes leading to the conversion of gas to stars need to be understood directly 
via observations of cold atomic and molecular gas. Indeed recent results using ALMA indicate that the cosmic molecular gas density peaks at $z\sim1.5$ and drops by a factor of $\sim6.5$ 
to $z\sim0$ \citep{decarli2019}, matching the evolution of the cosmic SFRD. Hence, the emerging picture is that cold gas content is the driving force behind the star formation history.

\begin{figure*}[t]
\includegraphics[width=0.4\textwidth, angle=90]{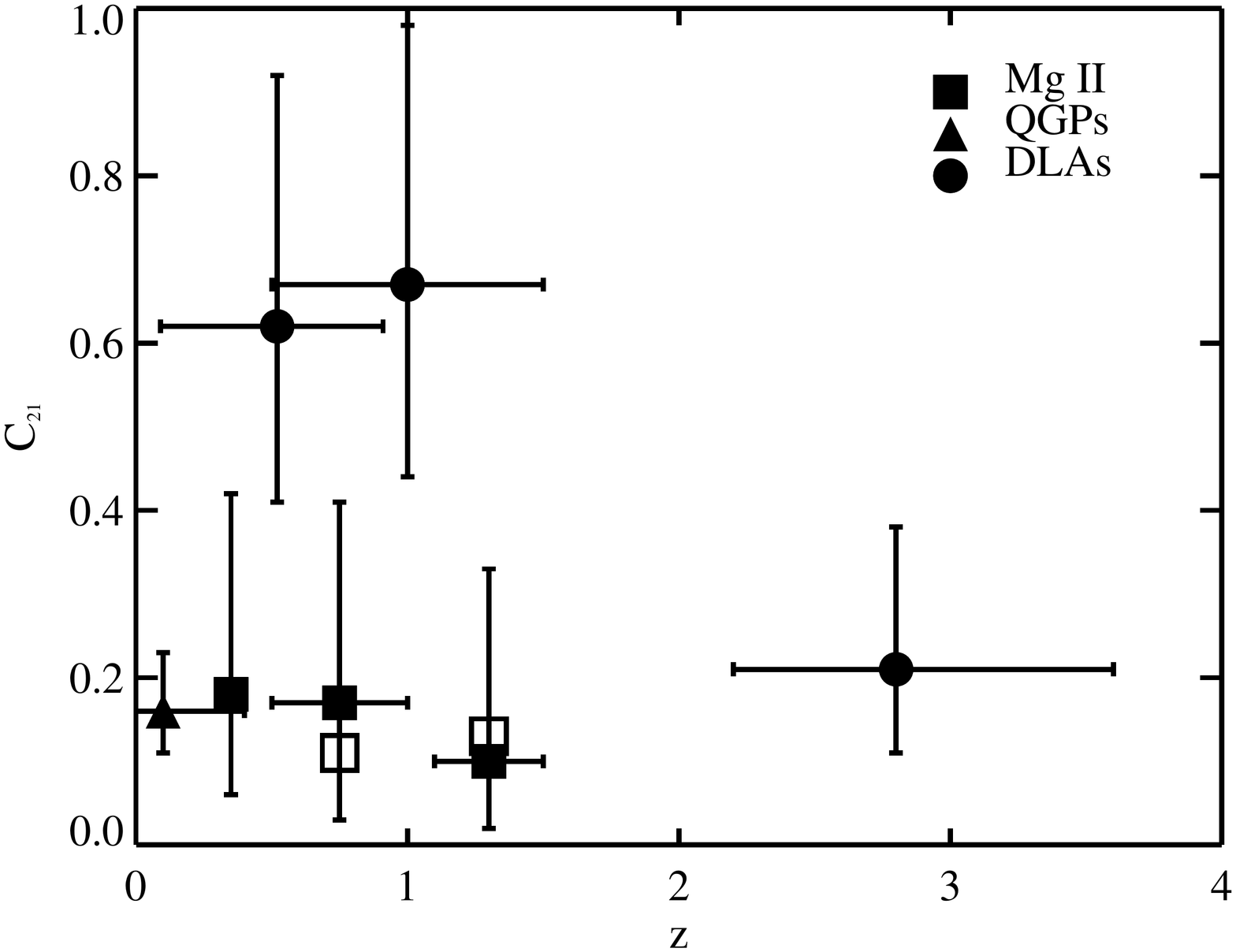}
\includegraphics[width=0.4\textwidth, angle=90]{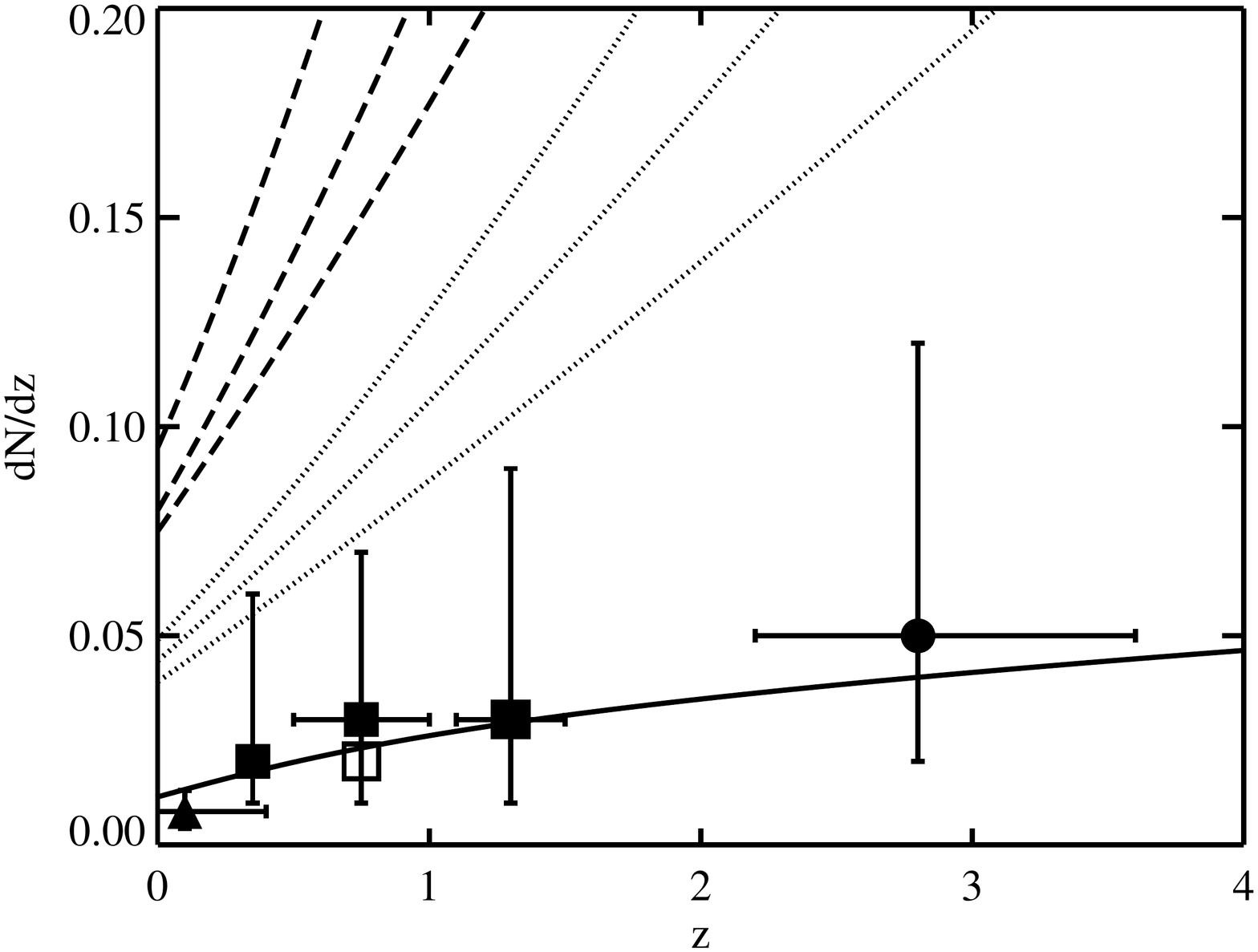}
\caption{Redshift evolution of detection rate (left) and number density per unit redshift (right) of \hi\ \21\ absorbers. The triangles, squares and circles represent estimates based 
on QGPs, strong \mgii\ systems and DLAs, respectively, from literature compilations \citep{lane2000,gupta2009,gupta2012,srianand2012,kanekar2009a,kanekar2014a,dutta2017a,dutta2017b,dutta2017c}. 
The open symbols represent estimates that have been corrected for partial coverage whenever possible \citep[see for details][]{gupta2012}. The dashed and dotted lines in the right panel 
shows for reference the redshift evolution of the number of strong \mgii\ absorbers (\wmg\ $\ge$ 1 \AA) per unit redshift \citep{prochter2006}, and the number of DLAs per unit redshift 
\citep{rao2006}, respectively. The solid line is the curve for non-evolving population of \hi\ \21\ absorbers normalized at $z$ = 1.3. {\it The next challenge for \hi\ \21\ absorption
line studies would be to reduce the large uncertainties, statistical and systematic, on these measurements, through blind unbiased searches for absorbers.}}
\label{fig:evolution}
\end{figure*}

\hi\ \21\ emission and absorption measurements have been used to study in detail the CNM phase in the ISM of the Milky Way \citep{heiles2003,roy2013}. It has been observed that $\sim$60\% 
of the atomic gas is in the WNM phase and $\sim$50\% of the WNM lies in the thermally unstable range of 500$-$5000 K. \citet{Jenkins2001} also arrived at similar conclusions based on excitation of neutral carbon using UV spectroscopy. The ideal, i.e. dust- and luminosity-unbiased, way to estimate the redshift evolution of the cold gas fraction 
would be to conduct blind \hi\ \21\ absorption searches. However, this has not been possible till recently due to limited receiver bandwidths and hostile radio frequency interference 
environment, though the situation is now improving with the advent of Square Kilometer Array (SKA) pre-cursors and induction of wide-band receivers in existing telescopes. Till date, 
\hi\ \21\ absorption studies have been typically conducted in samples of QGPs (see Section~\ref{sec:galaxy}) or absorption-selected samples (see Section~\ref{sec:absorption}). 
Fig.~\ref{fig:evolution} summarizes the detection rates of \hi\ \21\ absorption and the number density per unit redshift of \hi\ \21\ absorbers ($n_{21}$) obtained using different 
techniques at different redshifts. No significant evolution is observed in the detection rate and $n_{21}$ among strong \mgii\ systems over $0.3<z<1.5$. However, comparing \hi\ \21\ 
studies of $z>2$ and $z<1$ DLAs, it can be seen that the cold gas fraction in DLAs may be declining with redshift, with the detection rate being $\sim$3 times lower at $z>2$ compared to $z<1$.  

Besides the evolution in incidence, \citet{kanekar2014a} have found evidence for redshift evolution in DLA spin temperatures. The \ts\ distribution of $z>2$ DLAs is found to be significantly
different from that of $z<2$ DLAs, and the high temperatures in high-$z$ DLAs are attributed to lower fractions of CNM. Further, \citet{dutta2017b} have found an increasing trend of the velocity 
width of the \hi\ \21\ absorption lines with redshift. A possible explanation for this is that the typical \hi\ \21\ absorber may be probed by larger mass galaxy halos at high-$z$ (for a given 
metallicity). In addition, taking into account the evolution of size and luminosity of galaxies with redshift, the radius of the cold \hi\ gas around a galaxy that gives rise to \hi\ \21\ 
absorption is likely be much higher at high-$z$ than what is seen at low-$z$ for a galaxy with same optical luminosity. 

While we have made progress in understanding the nature of cold \hi\ gas in galaxies and its redshift evolution, it can be seen that we are still limited by statistical uncertainties due 
to the small number of detections and different sample selection techniques at different redshifts. Hence, increasing the number of \hi\ \21\ detections over $0<z<2$, such that we can 
uniformly trace the evolution of cold gas in galaxies in a dust- and luminosity-unbiased way, is one of the major motivations of the upcoming blind \hi\ \21\ absorption line surveys using 
the SKA pathfinders and pre-cursors, e.g. FLASH/ASKAP, MALS/MeerKAT, SHARP/Apertif \citep{gupta2016,maccagni2017}.

\section{Summary \& Future Perspectives}
\label{sec:conclusion}
We have discussed in this work the results from different efforts to study the cold neutral gas in and around galaxies using \hi\ \21\ absorption. Some of the key results from 
these studies are summarized here.
\begin{itemize}
 \item The cold neutral gas around $z<0.4$ galaxies, as traced by \hi\ \21\ absorption, has a weakly declining radial profile, with an average covering factor of $0.16^{+0.07}_{-0.05}$
       within 30 kpc. Based on geometrical and chemical analysis, the \hi\ \21\ absorbers are likely tracing the diffuse extended \hi\ disc around galaxies. The cold \hi\ gas is patchily
       distributed around galaxies and can have variations in the optical depth at both parsec- and kilo-parsec-scales. 
 \item The detection rate and optical depth of \hi\ \21\ absorption over $0.5<z<1.5$ are correlated with the equivalent width of metal line absorption like \mgii\ and \feii. They are 
       further correlated with the reddening of the background quasar. Thus, the presence and amount of cold \hi\ gas in the vicinity of high-$z$ galaxies appear to be closely related 
       to the metal and dust content of the gas.
 \item The incidence and number density per unit redshift of \hi\ \21\ absorbers in strong \mgii\ absorbers do not seem to evolve over $0.3<z<1.5$, albeit the uncertainties are still large.
       The incidence of \hi\ \21\ absorbers in DLAs, on the other hand, seem to show a stronger evolution, i.e. it increases by a factor of $\sim$3 from $z>2$ to $z<1$. The spin temperature
       of DLAs also show a redshift evolution, with $z>2$ DLAs having significantly higher \ts.
\end{itemize}

\hi\ \21\ absorption towards background radio-loud quasars has thus proved to be an effective tool to map the distribution of cold neutral gas around foreground low-$z$ galaxies, 
as well as study the neutral gas in the vicinity of high-$z$ galaxies selected via \mgii\ absorption or DLAs. The next step is to increase the number of sightlines searched for 
\hi\ \21\ absorption and the number of detections through blind surveys as mentioned in Section~\ref{sec:evolution} The large number ($\sim$100s) of \hi\ \21\ absorption detection 
is expected to accurately characterize the redshift evolution of cold gas in galaxies. Here we identify few prospects involving multi-wavelength observations that are promising 
for interpreting the existing and upcoming results with \hi\ \21\ absorption.

\begin{enumerate}
 \item High resolution ($R\sim40000$) optical spectroscopy of quasars which show \hi\ \21\ absorption will enable detailed comparison of the kinematics of metals and \hi\ gas, and confirm
       with higher significance the various trends of \hi\ with metals and dust as found using low resolution optical spectra \citep[e.g.][]{srianand2012}. 
 \item Sub-arcsecond-scale imaging of the background radio sources will allow us to quantify the covering factor of cold gas and study its small scale structure \citep[e.g.][]{gupta2012}. 
 \item OH and CO absorption spectroscopy of \hi\ \21\ absorbers will facilitate comparison of the kinematics and structure of neutral and molecular gas phases in and around galaxies 
       \citep[e.g.][]{gupta2018b,combes2019}. 
 \item Identifying the host galaxies of high-$z$ \hi\ \21\ absorbers using optical integral field unit spectroscopy (e.g. MUSE/VLT) and molecular emission with e.g. ALMA, is essential 
       to connect the properties of the galaxies with the absorbing gas \citep[e.g.][]{peroux2019}. Such observations are also required to link the neutral gas around galaxies with their 
       ionized and molecular gas content, and hence understand the mechanisms through which galaxies acquire the fuel for forming stars. 
\end{enumerate}

The upcoming surveys will provide interesting targets for follow-up multi-wavelength observations as outlined above. In addition, existing and planned wide-area multi-object/integral field 
optical spectroscopic surveys like Calar Alto Legacy Integral Field Area \citep[CALIFA;][]{sanchez2012}, Sydney-Australian-Astronomical-Observatory Multiobject Integral-Field Spectrograph 
\citep[SAMI;][]{croom2012}, SDSS Mapping Nearby Galaxies at Apache Point Observatory \citep[MaNGA;][]{bundy2015} and WEAVE \citep{dalton2012}, will complement the \hi\ \21\ surveys in 
associating the absorption with the galaxies.

\section*{Acknowledgements}
I thank the anonymous reviewers for their helpful comments.
I thank Raghunathan Srianand for going through a draft of this review and providing feedback.
I am grateful to the Alexander von Humboldt Foundation for support in the form of a post-doctoral fellowship.


%
\def\aap{A\&A}%
\def\aapr{A\&A~Rev.}%
\def\aaps{A\&AS}%
\def\aj{AJ}%
\def\actaa{Acta Astron.}%
\def\araa{ARA\&A}%
\def\apj{ApJ}%
\def\apjl{ApJ}%
\def\apjs{ApJS}%
\def\apspr{Astrophys.~Space~Phys.~Res.}%
\def\ao{Appl.~Opt.}%
\def\aplett{Astrophys.~Lett.}%
\def\apss{Ap\&SS}%
\def\azh{AZh}%
\def\bain{Bull.~Astron.~Inst.~Netherlands}%
\def\baas{BAAS}%
\def\bac{Bull. astr. Inst. Czechosl.}%
\def\caa{Chinese Astron. Astrophys.}%
\def\cjaa{Chinese J. Astron. Astrophys.}%
\def\fcp{Fund.~Cosmic~Phys.}%
\def\gafd{Geophys.\ Astrophys.\ Fluid Dyn.}
\def\gca{Geochim.~Cosmochim.~Acta}%
\def\grl{Geophys.~Res.~Lett.}%
\def\iaucirc{IAU~Circ.}%
\def\icarus{Icarus}%
\def\jcap{J. Cosmology Astropart. Phys.}%
\def\jcp{J.~Chem.~Phys.}%
\def\jfm{JFM}
\def\jgr{J.~Geophys.~Res.}%
\def\jqsrt{J.~Quant.~Spec.~Radiat.~Transf.}%
\def\jrasc{JRASC}%
\def\mnras{MNRAS}%
\def\memras{MmRAS}%
\def\memsai{Mem.~Soc.~Astron.~Italiana}%
\def\na{New A}%
\def\nar{New A Rev.}%
\def\nat{Nature}%
\def\nphysa{Nucl.~Phys.~A}%
\def\pasa{PASA}%
\def\pasj{PASJ}%
\def\pasp{PASP}%
\def\physrep{Phys.~Rep.}%
\def\physscr{Phys.~Scr}%
\def\planss{Planet.~Space~Sci.}%
\def\pra{Phys.~Rev.~A}%
\def\prb{Phys.~Rev.~B}%
\def\prc{Phys.~Rev.~C}%
\def\prd{Phys.~Rev.~D}%
\def\pre{Phys.~Rev.~E}%
\def\prl{Phys.~Rev.~Lett.}%
\def\procspie{Proc.~SPIE}%
\def\qjras{QJRAS}%
\def\rmxaa{Rev. Mexicana Astron. Astrofis.}%
\def\sgg{Stud.\ Geoph.\ et\ Geod.}
\def\skytel{S\&T}%
\def\solphys{Sol.~Phys.}%
\def\sovast{Soviet~Ast.}%
\def\ssr{Space~Sci.~Rev.}%
\def\zap{ZAp}%
\def\memsai{Memorie della Societa Astronomica Italiana}
\def\araa{Ann. Rev. of Astron. and Astrophys.}
\newcommand\zhetp{JETP}
\newcommand\jetp{Sov.\ Phys.\ JETP}
\let\astap=\aap
\let\apjlett=\apjl
\let\apjsupp=\apjs
\let\applopt=\ao
\bibliographystyle{spmpsci}
\bibliography{mybib}

\begin{thebibliography}{}
\expandafter\ifx\csname natexlab\endcsname\relax\def\natexlab#1{#1}\fi

\bibitem[{{Bagetakos} {et~al.}(2011){Bagetakos}, {Brinks}, {Walter}, {de Blok},
  {Usero}, {Leroy}, {Rich}, \& {Kennicutt}}]{bagetakos2011}
{Bagetakos}, I., {Brinks}, E., {Walter}, F., {et~al.} 2011, \aj, 141, 23

\bibitem[{{Bahcall} \& {Ekers}(1969)}]{bahcall1969}
{Bahcall}, J.~N., \& {Ekers}, R.~D. 1969, \apj, 157, 1055

\bibitem[{{Battisti} {et~al.}(2012){Battisti}, {Meiring}, {Tripp}, {Prochaska},
  {Werk}, {Jenkins}, {Lehner}, {Tumlinson}, \& {Thom}}]{battisti2012}
{Battisti}, A.~J., {Meiring}, J.~D., {Tripp}, T.~M., {et~al.} 2012, \apj, 744,
  93

\bibitem[{{Begeman}(1987)}]{begeman1987}
{Begeman}, K.~G. 1987, PhD thesis, , Kapteyn Institute, (1987)

\bibitem[{{Begum} {et~al.}(2008){Begum}, {Chengalur}, {Karachentsev},
  {Sharina}, \& {Kaisin}}]{begum2008}
{Begum}, A., {Chengalur}, J.~N., {Karachentsev}, I.~D., {Sharina}, M.~E., \&
  {Kaisin}, S.~S. 2008, \mnras, 386, 1667

\bibitem[{{Bergeron} \& {Boiss{\'e}}(1991)}]{bergeron1991}
{Bergeron}, J., \& {Boiss{\'e}}, P. 1991, \aap, 243, 344

\bibitem[{{Biggs} {et~al.}(2016){Biggs}, {Zwaan}, {Hatziminaoglou},
  {P{\'e}roux}, \& {Liske}}]{biggs2016}
{Biggs}, A.~D., {Zwaan}, M.~A., {Hatziminaoglou}, E., {P{\'e}roux}, C., \&
  {Liske}, J. 2016, \mnras, 462, 2819

\bibitem[{{Bordoloi} {et~al.}(2014{\natexlab{a}}){Bordoloi}, {Lilly},
  {Kacprzak}, \& {Churchill}}]{bordoloi2014a}
{Bordoloi}, R., {Lilly}, S.~J., {Kacprzak}, G.~G., \& {Churchill}, C.~W.
  2014{\natexlab{a}}, \apj, 784, 108

\bibitem[{{Bordoloi} {et~al.}(2014{\natexlab{b}}){Bordoloi}, {Tumlinson},
  {Werk}, {Oppenheimer}, {Peeples}, {Prochaska}, {Tripp}, {Katz}, {Dav{\'e}},
  {Fox}, {Thom}, {Ford}, {Weinberg}, {Burchett}, \&
  {Kollmeier}}]{bordoloi2014b}
{Bordoloi}, R., {Tumlinson}, J., {Werk}, J.~K., {et~al.} 2014{\natexlab{b}},
  \apj, 796, 136

\bibitem[{{Borthakur}(2016)}]{borthakur2016}
{Borthakur}, S. 2016, \apj, 829, 128

\bibitem[{{Borthakur} {et~al.}(2011){Borthakur}, {Tripp}, {Yun}, {Bowen},
  {Meiring}, {York}, \& {Momjian}}]{borthakur2011}
{Borthakur}, S., {Tripp}, T.~M., {Yun}, M.~S., {et~al.} 2011, \apj, 727, 52

\bibitem[{{Borthakur} {et~al.}(2015){Borthakur}, {Heckman}, {Tumlinson},
  {Bordoloi}, {Thom}, {Catinella}, {Schiminovich}, {Dav{\'e}}, {Kauffmann},
  {Moran}, \& {Saintonge}}]{borthakur2015}
{Borthakur}, S., {Heckman}, T., {Tumlinson}, J., {et~al.} 2015, \apj, 813, 46

\bibitem[{{Bosma}(1978)}]{bosma1978}
{Bosma}, A. 1978, PhD thesis, PhD Thesis, Groningen Univ., (1978)

\bibitem[{{Bosma}(1981{\natexlab{a}})}]{bosma1981a}
---. 1981{\natexlab{a}}, \aj, 86, 1791

\bibitem[{{Bosma}(1981{\natexlab{b}})}]{bosma1981b}
---. 1981{\natexlab{b}}, \aj, 86, 1825

\bibitem[{{Braun}(1997)}]{braun1997}
{Braun}, R. 1997, \apj, 484, 637

\bibitem[{{Briggs} \& {Wolfe}(1983)}]{briggs1983}
{Briggs}, F.~H., \& {Wolfe}, A.~M. 1983, \apj, 268, 76

\bibitem[{{Broeils}(1992)}]{broeils1992}
{Broeils}, A.~H. 1992, PhD thesis, PhD thesis, Univ.~Groningen, (1992)

\bibitem[{{Brown} \& {Roberts}(1973)}]{brown1973}
{Brown}, R.~L., \& {Roberts}, M.~S. 1973, \apjl, 184, L7

\bibitem[{{Bundy} {et~al.}(2015){Bundy}, {Bershady}, {Law}, {Yan}, {Drory},
  {MacDonald}, {Wake}, {Cherinka}, {S{\'a}nchez-Gallego}, {Weijmans}, {Thomas},
  {Tremonti}, {Masters}, {Coccato}, {Diamond-Stanic}, {Arag{\'o}n-Salamanca},
  {Avila-Reese}, {Badenes}, {Falc{\'o}n-Barroso}, {Belfiore}, {Bizyaev},
  {Blanc}, {Bland-Hawthorn}, {Blanton}, {Brownstein}, {Byler}, {Cappellari},
  {Conroy}, {Dutton}, {Emsellem}, {Etherington}, {Frinchaboy}, {Fu}, {Gunn},
  {Harding}, {Johnston}, {Kauffmann}, {Kinemuchi}, {Klaene}, {Knapen},
  {Leauthaud}, {Li}, {Lin}, {Maiolino}, {Malanushenko}, {Malanushenko}, {Mao},
  {Maraston}, {McDermid}, {Merrifield}, {Nichol}, {Oravetz}, {Pan}, {Parejko},
  {Sanchez}, {Schlegel}, {Simmons}, {Steele}, {Steinmetz}, {Thanjavur},
  {Thompson}, {Tinker}, {van den Bosch}, {Westfall}, {Wilkinson}, {Wright},
  {Xiao}, \& {Zhang}}]{bundy2015}
{Bundy}, K., {Bershady}, M.~A., {Law}, D.~R., {et~al.} 2015, \apj, 798, 7

\bibitem[{{Carilli} {et~al.}(1998){Carilli}, {Menten}, {Reid}, {Rupen}, \&
  {Yun}}]{carilli1998}
{Carilli}, C.~L., {Menten}, K.~M., {Reid}, M.~J., {Rupen}, M.~P., \& {Yun},
  M.~S. 1998, \apj, 494, 175

\bibitem[{{Carilli} {et~al.}(1992){Carilli}, {Perlman}, \&
  {Stocke}}]{carilli1992b}
{Carilli}, C.~L., {Perlman}, E.~S., \& {Stocke}, J.~T. 1992, \apjl, 400, L13

\bibitem[{{Carilli} \& {van Gorkom}(1992)}]{carilli1992a}
{Carilli}, C.~L., \& {van Gorkom}, J.~H. 1992, \apj, 399, 373

\bibitem[{{Carilli} {et~al.}(2000){Carilli}, {Menten}, {Stocke}, {Perlman},
  {Vermeulen}, {Briggs}, {de Bruyn}, {Conway}, \& {Moore}}]{carilli2000}
{Carilli}, C.~L., {Menten}, K.~M., {Stocke}, J.~T., {et~al.} 2000, \prl, 85,
  5511

\bibitem[{{Catinella} \& {Cortese}(2015)}]{catinella2015}
{Catinella}, B., \& {Cortese}, L. 2015, \mnras, 446, 3526

\bibitem[{{Chang} {et~al.}(2010){Chang}, {Pen}, {Bandura}, \&
  {Peterson}}]{chang2010}
{Chang}, T.-C., {Pen}, U.-L., {Bandura}, K., \& {Peterson}, J.~B. 2010, \nat,
  466, 463

\bibitem[{{Chen} {et~al.}(2010){Chen}, {Helsby}, {Gauthier}, {Shectman},
  {Thompson}, \& {Tinker}}]{chen2010a}
{Chen}, H.-W., {Helsby}, J.~E., {Gauthier}, J.-R., {et~al.} 2010, \apj, 714,
  1521

\bibitem[{{Chen} {et~al.}(2001){Chen}, {Lanzetta}, {Webb}, \&
  {Barcons}}]{chen2001}
{Chen}, H.-W., {Lanzetta}, K.~M., {Webb}, J.~K., \& {Barcons}, X. 2001, \apj,
  559, 654

\bibitem[{{Chengalur} {et~al.}(1999){Chengalur}, {de Bruyn}, \&
  {Narasimha}}]{chengalur1999}
{Chengalur}, J.~N., {de Bruyn}, A.~G., \& {Narasimha}, D. 1999, \aap, 343, L79

\bibitem[{{Chengalur} \& {Kanekar}(2003)}]{chengalur2003}
{Chengalur}, J.~N., \& {Kanekar}, N. 2003, \prl, 91, 241302

\bibitem[{{Chung} {et~al.}(2009){Chung}, {van Gorkom}, {Kenney}, {Crowl}, \&
  {Vollmer}}]{chung2009}
{Chung}, A., {van Gorkom}, J.~H., {Kenney}, J.~D.~P., {Crowl}, H., \&
  {Vollmer}, B. 2009, \aj, 138, 1741

\bibitem[{{Churchill} {et~al.}(2013){Churchill}, {Nielsen}, {Kacprzak}, \&
  {Trujillo-Gomez}}]{churchill2013}
{Churchill}, C.~W., {Nielsen}, N.~M., {Kacprzak}, G.~G., \& {Trujillo-Gomez},
  S. 2013, \apjl, 763, L42

\bibitem[{{Clark} {et~al.}(1962){Clark}, {Radhakrishnan}, \&
  {Wilson}}]{clark1962}
{Clark}, B.~G., {Radhakrishnan}, V., \& {Wilson}, R.~W. 1962, \apj, 135, 151

\bibitem[{{Combes} {et~al.}(2019){Combes}, {Gupta}, {Jozsa}, \&
  {Momjian}}]{combes2019}
{Combes}, F., {Gupta}, N., {Jozsa}, G.~I.~G., \& {Momjian}, E. 2019, \aap, 623,
  A133

\bibitem[{{Croom} {et~al.}(2012){Croom}, {Lawrence}, {Bland-Hawthorn},
  {Bryant}, {Fogarty}, {Richards}, {Goodwin}, {Farrell}, {Miziarski}, {Heald},
  {Jones}, {Lee}, {Colless}, {Brough}, {Hopkins}, {Bauer}, {Birchall}, {Ellis},
  {Horton}, {Leon-Saval}, {Lewis}, {L{\'o}pez-S{\'a}nchez}, {Min}, {Trinh}, \&
  {Trowland}}]{croom2012}
{Croom}, S.~M., {Lawrence}, J.~S., {Bland-Hawthorn}, J., {et~al.} 2012, \mnras,
  421, 872

\bibitem[{{Curran} {et~al.}(2005){Curran}, {Murphy}, {Pihlstr{\"o}m}, {Webb},
  \& {Purcell}}]{curran2005}
{Curran}, S.~J., {Murphy}, M.~T., {Pihlstr{\"o}m}, Y.~M., {Webb}, J.~K., \&
  {Purcell}, C.~R. 2005, \mnras, 356, 1509

\bibitem[{{Curran} {et~al.}(2010){Curran}, {Tzanavaris}, {Darling}, {Whiting},
  {Webb}, {Bignell}, {Athreya}, \& {Murphy}}]{curran2010a}
{Curran}, S.~J., {Tzanavaris}, P., {Darling}, J.~K., {et~al.} 2010, \mnras,
  402, 35

\bibitem[{{Dalton} {et~al.}(2012){Dalton}, {Trager}, {Abrams}, {Carter},
  {Bonifacio}, {Aguerri}, {MacIntosh}, {Evans}, {Lewis}, {Navarro}, {Agocs},
  {Dee}, {Rousset}, {Tosh}, {Middleton}, {Pragt}, {Terrett}, {Brock}, {Benn},
  {Verheijen}, {Cano Infantes}, {Bevil}, {Steele}, {Mottram}, {Bates},
  {Gribbin}, {Rey}, {Rodriguez}, {Delgado}, {Guinouard}, {Walton}, {Irwin},
  {Jagourel}, {Stuik}, {Gerlofsma}, {Roelfsma}, {Skillen}, {Ridings},
  {Balcells}, {Daban}, {Gouvret}, {Venema}, \& {Girard}}]{dalton2012}
{Dalton}, G., {Trager}, S.~C., {Abrams}, D.~C., {et~al.} 2012, in \procspie,
  Vol. 8446, Ground-based and Airborne Instrumentation for Astronomy IV, 84460P

\bibitem[{{de Blok} {et~al.}(2008){de Blok}, {Walter}, {Brinks},
  {Trachternach}, {Oh}, \& {Kennicutt}}]{deblok2008}
{de Blok}, W.~J.~G., {Walter}, F., {Brinks}, E., {et~al.} 2008, \aj, 136, 2648

\bibitem[{{Decarli} {et~al.}(2019){Decarli}, {Walter},
  {G{\'o}nzalez-L{\'o}pez}, {Aravena}, {Boogaard}, {Carilli}, {Cox}, {Daddi},
  {Popping}, {Riechers}, {Uzgil}, {Weiss}, {Assef}, {Bacon}, {Bauer},
  {Bertoldi}, {Bouwens}, {Contini}, {Cortes}, {da Cunha}, {D{\'{\i}}az-Santos},
  {Elbaz}, {Inami}, {Hodge}, {Ivison}, {Le F{\`e}vre}, {Magnelli}, {Novak},
  {Oesch}, {Rix}, {Sargent}, {Smail}, {Swinbank}, {Somerville}, {van der Werf},
  {Wagg}, \& {Wisotzki}}]{decarli2019}
{Decarli}, R., {Walter}, F., {G{\'o}nzalez-L{\'o}pez}, J., {et~al.} 2019, arXiv
  e-prints, arXiv:1903.09164

\bibitem[{{Delhaize} {et~al.}(2013){Delhaize}, {Meyer}, {Staveley-Smith}, \&
  {Boyle}}]{delhaize2013}
{Delhaize}, J., {Meyer}, M.~J., {Staveley-Smith}, L., \& {Boyle}, B.~J. 2013,
  \mnras, 433, 1398

\bibitem[{{Dickey} {et~al.}(1983){Dickey}, {Kulkarni}, {van Gorkom}, \&
  {Heiles}}]{dickey1983}
{Dickey}, J.~M., {Kulkarni}, S.~R., {van Gorkom}, J.~H., \& {Heiles}, C.~E.
  1983, \apjs, 53, 591

\bibitem[{{Dickey} {et~al.}(2000){Dickey}, {Mebold}, {Stanimirovic}, \&
  {Staveley-Smith}}]{dickey2000}
{Dickey}, J.~M., {Mebold}, U., {Stanimirovic}, S., \& {Staveley-Smith}, L.
  2000, \apj, 536, 756

\bibitem[{{Dutta} {et~al.}(2016){Dutta}, {Gupta}, {Srianand}, \&
  {O'Meara}}]{dutta2016}
{Dutta}, R., {Gupta}, N., {Srianand}, R., \& {O'Meara}, J.~M. 2016, \mnras,
  456, 4209

\bibitem[{{Dutta} {et~al.}(2017{\natexlab{a}}){Dutta}, {Srianand}, {Gupta}, \&
  {Joshi}}]{dutta2017c}
{Dutta}, R., {Srianand}, R., {Gupta}, N., \& {Joshi}, R. 2017{\natexlab{a}},
  ArXiv e-prints, arXiv:1703.00457

\bibitem[{{Dutta} {et~al.}(2017{\natexlab{b}}){Dutta}, {Srianand}, {Gupta},
  {Joshi}, {Petitjean}, {Noterdaeme}, {Ge}, \& {Krogager}}]{dutta2017b}
{Dutta}, R., {Srianand}, R., {Gupta}, N., {et~al.} 2017{\natexlab{b}}, \mnras,
  465, 4249

\bibitem[{{Dutta} {et~al.}(2017{\natexlab{c}}){Dutta}, {Srianand}, {Gupta},
  {Momjian}, {Noterdaeme}, {Petitjean}, \& {Rahmani}}]{dutta2017a}
---. 2017{\natexlab{c}}, \mnras, 465, 588

\bibitem[{{Dutta} {et~al.}(2015){Dutta}, {Srianand}, {Muzahid}, {Gupta},
  {Momjian}, \& {Charlton}}]{dutta2015}
{Dutta}, R., {Srianand}, R., {Muzahid}, S., {et~al.} 2015, \mnras, 448, 3718

\bibitem[{{Ewen} \& {Purcell}(1951)}]{ewen1951}
{Ewen}, H.~I., \& {Purcell}, E.~M. 1951, \nat, 168, 356

\bibitem[{{Fern{\'a}ndez} {et~al.}(2016){Fern{\'a}ndez}, {Gim}, {van Gorkom},
  {Yun}, {Momjian}, {Popping}, {Chomiuk}, {Hess}, {Hunt}, {Kreckel}, {Lucero},
  {Maddox}, {Oosterloo}, {Pisano}, {Verheijen}, {Hales}, {Chung}, {Dodson},
  {Golap}, {Gross}, {Henning}, {Hibbard}, {Jaff{\'e}}, {Donovan Meyer},
  {Meyer}, {Sanchez-Barrantes}, {Schiminovich}, {Wicenec}, {Wilcots},
  {Bershady}, {Scoville}, {Strader}, {Tremou}, {Salinas}, \&
  {Ch{\'a}vez}}]{fernandez2016}
{Fern{\'a}ndez}, X., {Gim}, H.~B., {van Gorkom}, J.~H., {et~al.} 2016, \apjl,
  824, L1

\bibitem[{{Field}(1959)}]{field1959}
{Field}, G.~B. 1959, \apj, 129, 536

\bibitem[{{Field} {et~al.}(1969){Field}, {Goldsmith}, \& {Habing}}]{field1969}
{Field}, G.~B., {Goldsmith}, D.~W., \& {Habing}, H.~J. 1969, \apjl, 155, L149

\bibitem[{{Fynbo} {et~al.}(2008){Fynbo}, {Prochaska}, {Sommer-Larsen},
  {Dessauges-Zavadsky}, \& {M{\o}ller}}]{fynbo2008}
{Fynbo}, J.~P.~U., {Prochaska}, J.~X., {Sommer-Larsen}, J.,
  {Dessauges-Zavadsky}, M., \& {M{\o}ller}, P. 2008, \apj, 683, 321

\bibitem[{{Gupta} {et~al.}(2018{\natexlab{a}}){Gupta}, {Momjian}, {Srianand},
  {Petitjean}, {Noterdaeme}, {Gyanchandani}, {Sharma}, \&
  {Kulkarni}}]{gupta2018b}
{Gupta}, N., {Momjian}, E., {Srianand}, R., {et~al.} 2018{\natexlab{a}}, \apjl,
  860, L22

\bibitem[{{Gupta} {et~al.}(2010){Gupta}, {Srianand}, {Bowen}, {York}, \&
  {Wadadekar}}]{gupta2010}
{Gupta}, N., {Srianand}, R., {Bowen}, D.~V., {York}, D.~G., \& {Wadadekar}, Y.
  2010, \mnras, 408, 849

\bibitem[{{Gupta} {et~al.}(2012){Gupta}, {Srianand}, {Petitjean}, {Bergeron},
  {Noterdaeme}, \& {Muzahid}}]{gupta2012}
{Gupta}, N., {Srianand}, R., {Petitjean}, P., {et~al.} 2012, \aap, 544, A21

\bibitem[{{Gupta} {et~al.}(2009){Gupta}, {Srianand}, {Petitjean}, {Noterdaeme},
  \& {Saikia}}]{gupta2009}
{Gupta}, N., {Srianand}, R., {Petitjean}, P., {Noterdaeme}, P., \& {Saikia},
  D.~J. 2009, \mnras, 398, 201

\bibitem[{{Gupta} {et~al.}(2016){Gupta}, {Srianand}, {Baan}, {Baker},
  {Beswick}, {Bhatnagar}, {Bhattacharya}, {Bosma}, {Carilli}, {Cluver},
  {Combes}, {Cress}, {Dutta}, {Fynbo}, {Heald}, {Hilton}, {Hussain}, {Jarvis},
  {Jozsa}, {Kamphuis}, {Kembhavi}, {Kerp}, {Kloeckner}, {Krogager}, {Kulkarni},
  {Ledoux}, {Mahabal}, {Mauch}, {Moodley}, {Momjian}, {Morganti}, {Noterdaeme},
  {Oosterloo}, {Petitjean}, {Schroeder}, {Serra}, {Sievers}, {Spekkens},
  {Vaisanen}, {van der Hulst}, {Vivek}, {Wang}, {Wong}, \& {Zungu}}]{gupta2016}
{Gupta}, N., {Srianand}, R., {Baan}, W., {et~al.} 2016, in Proceedings of
  MeerKAT Science: On the Pathway to the SKA. 25-27 May, 2016 Stellenbosch,
  South Africa (MeerKAT2016)., 14

\bibitem[{{Gupta} {et~al.}(2018{\natexlab{b}}){Gupta}, {Srianand}, {Farnes},
  {Pidopryhora}, {Vivek}, {Paragi}, {Noterdaeme}, {Oosterloo}, \&
  {Petitjean}}]{gupta2018a}
{Gupta}, N., {Srianand}, R., {Farnes}, J.~S., {et~al.} 2018{\natexlab{b}},
  \mnras, 476, 2432

\bibitem[{{Hagen} {et~al.}(1955){Hagen}, {Lilley}, \& {McClain}}]{hagen1955}
{Hagen}, J.~P., {Lilley}, A.~E., \& {McClain}, E.~F. 1955, \apj, 122, 361

\bibitem[{{Hagen} \& {McClain}(1954)}]{hagen1954b}
{Hagen}, J.~P., \& {McClain}, E.~F. 1954, \apj, 120, 368

\bibitem[{{Hagen} {et~al.}(1954){Hagen}, {McClain}, \& {Hepburn}}]{hagen1954a}
{Hagen}, J.~P., {McClain}, E.~F., \& {Hepburn}, N. 1954, \aj, 59, 323

\bibitem[{{Haynes} {et~al.}(1979){Haynes}, {Giovanelli}, \&
  {Roberts}}]{haynes1979}
{Haynes}, M.~P., {Giovanelli}, R., \& {Roberts}, M.~S. 1979, \apj, 229, 83

\bibitem[{{Heiles} \& {Miley}(1970)}]{heiles1970}
{Heiles}, C., \& {Miley}, G.~K. 1970, \apjl, 160, L83

\bibitem[{{Heiles} \& {Troland}(2003)}]{heiles2003}
{Heiles}, C., \& {Troland}, T.~H. 2003, \apj, 586, 1067

\bibitem[{{Heiles} \& {Troland}(2004)}]{heiles2004}
---. 2004, \apjs, 151, 271

\bibitem[{{Hopkins} \& {Beacom}(2006)}]{hopkins2006}
{Hopkins}, A.~M., \& {Beacom}, J.~F. 2006, \apj, 651, 142

\bibitem[{{Hoppmann} {et~al.}(2015){Hoppmann}, {Staveley-Smith}, {Freudling},
  {Zwaan}, {Minchin}, \& {Calabretta}}]{hoppmann2015}
{Hoppmann}, L., {Staveley-Smith}, L., {Freudling}, W., {et~al.} 2015, \mnras,
  452, 3726

\bibitem[{{Ianjamasimanana} {et~al.}(2012){Ianjamasimanana}, {de Blok},
  {Walter}, \& {Heald}}]{ianjamasimanana2012}
{Ianjamasimanana}, R., {de Blok}, W.~J.~G., {Walter}, F., \& {Heald}, G.~H.
  2012, \aj, 144, 96

\bibitem[{{Ishwara-Chandra} {et~al.}(2003){Ishwara-Chandra}, {Dwarakanath}, \&
  {Anantharamaiah}}]{ishwara2003}
{Ishwara-Chandra}, C.~H., {Dwarakanath}, K.~S., \& {Anantharamaiah}, K.~R.
  2003, Journal of Astrophysics and Astronomy, 24, 37

\bibitem[{{Jenkins} \& {Tripp}(2001)}]{Jenkins2001}
{Jenkins}, E.~B., \& {Tripp}, T.~M. 2001, \apjs, 137, 297

\bibitem[{{Jones} {et~al.}(2018){Jones}, {Haynes}, {Giovanelli}, \&
  {Moorman}}]{jones2018}
{Jones}, M.~G., {Haynes}, M.~P., {Giovanelli}, R., \& {Moorman}, C. 2018,
  \mnras, 477, 2

\bibitem[{{Kacprzak} {et~al.}(2012){Kacprzak}, {Churchill}, \&
  {Nielsen}}]{kacprzak2012}
{Kacprzak}, G.~G., {Churchill}, C.~W., \& {Nielsen}, N.~M. 2012, \apjl, 760, L7

\bibitem[{{Kanekar} \& {Briggs}(2003)}]{kanekar2003b}
{Kanekar}, N., \& {Briggs}, F.~H. 2003, \aap, 412, L29

\bibitem[{{Kanekar} \& {Chengalur}(2003)}]{kanekar2003a}
{Kanekar}, N., \& {Chengalur}, J.~N. 2003, \aap, 399, 857

\bibitem[{{Kanekar} {et~al.}(2010){Kanekar}, {Chengalur}, \&
  {Ghosh}}]{kanekar2010}
{Kanekar}, N., {Chengalur}, J.~N., \& {Ghosh}, T. 2010, \apjl, 716, L23

\bibitem[{{Kanekar} {et~al.}(2001){Kanekar}, {Chengalur}, {Subrahmanyan}, \&
  {Petitjean}}]{kanekar2001}
{Kanekar}, N., {Chengalur}, J.~N., {Subrahmanyan}, R., \& {Petitjean}, P. 2001,
  \aap, 367, 46

\bibitem[{{Kanekar} {et~al.}(2013){Kanekar}, {Ellison}, {Momjian}, {York}, \&
  {Pettini}}]{kanekar2013}
{Kanekar}, N., {Ellison}, S.~L., {Momjian}, E., {York}, B.~A., \& {Pettini}, M.
  2013, \mnras, 428, 532

\bibitem[{{Kanekar} {et~al.}(2009{\natexlab{a}}){Kanekar}, {Lane}, {Momjian},
  {Briggs}, \& {Chengalur}}]{kanekar2009b}
{Kanekar}, N., {Lane}, W.~M., {Momjian}, E., {Briggs}, F.~H., \& {Chengalur},
  J.~N. 2009{\natexlab{a}}, \mnras, 394, L61

\bibitem[{{Kanekar} {et~al.}(2012){Kanekar}, {Langston}, {Stocke}, {Carilli},
  \& {Menten}}]{kanekar2012}
{Kanekar}, N., {Langston}, G.~I., {Stocke}, J.~T., {Carilli}, C.~L., \&
  {Menten}, K.~M. 2012, \apjl, 746, L16

\bibitem[{{Kanekar} {et~al.}(2009{\natexlab{b}}){Kanekar}, {Prochaska},
  {Ellison}, \& {Chengalur}}]{kanekar2009a}
{Kanekar}, N., {Prochaska}, J.~X., {Ellison}, S.~L., \& {Chengalur}, J.~N.
  2009{\natexlab{b}}, \mnras, 396, 385

\bibitem[{{Kanekar} {et~al.}(2016){Kanekar}, {Sethi}, \&
  {Dwarakanath}}]{kanekar2016}
{Kanekar}, N., {Sethi}, S., \& {Dwarakanath}, K.~S. 2016, \apjl, 818, L28

\bibitem[{{Kanekar} {et~al.}(2014){Kanekar}, {Prochaska}, {Smette}, {Ellison},
  {Ryan-Weber}, {Momjian}, {Briggs}, {Lane}, {Chengalur}, {Delafosse}, {Grave},
  {Jacobsen}, \& {de Bruyn}}]{kanekar2014a}
{Kanekar}, N., {Prochaska}, J.~X., {Smette}, A., {et~al.} 2014, \mnras, 438,
  2131

\bibitem[{{Keeney} {et~al.}(2011){Keeney}, {Stocke}, {Danforth}, \&
  {Carilli}}]{keeney2011}
{Keeney}, B.~A., {Stocke}, J.~T., {Danforth}, C.~W., \& {Carilli}, C.~L. 2011,
  \aj, 141, 66

\bibitem[{{Krogager} {et~al.}(2012){Krogager}, {Fynbo}, {M{\o}ller}, {Ledoux},
  {Noterdaeme}, {Christensen}, {Milvang-Jensen}, \& {Sparre}}]{krogager2012}
{Krogager}, J.-K., {Fynbo}, J.~P.~U., {M{\o}ller}, P., {et~al.} 2012, \mnras,
  424, L1

\bibitem[{{Kulkarni} \& {Heiles}(1987)}]{kulkarni1987}
{Kulkarni}, S.~R., \& {Heiles}, C. 1987, in Astrophysics and Space Science
  Library, Vol. 134, Interstellar Processes, ed. D.~J. {Hollenbach} \& H.~A.
  {Thronson}, Jr., 87--122

\bibitem[{{Kulkarni} \& {Heiles}(1988)}]{kulkarni1988}
{Kulkarni}, S.~R., \& {Heiles}, C. 1988, {Neutral hydrogen and the diffuse
  interstellar medium}, ed. K.~I. {Kellermann} \& G.~L. {Verschuur}, 95--153

\bibitem[{{Lah} {et~al.}(2007){Lah}, {Chengalur}, {Briggs}, {Colless}, {de
  Propris}, {Pracy}, {de Blok}, {Fujita}, {Ajiki}, {Shioya}, {Nagao},
  {Murayama}, {Taniguchi}, {Yagi}, \& {Okamura}}]{lah2007}
{Lah}, P., {Chengalur}, J.~N., {Briggs}, F.~H., {et~al.} 2007, \mnras, 376,
  1357

\bibitem[{{Lah} {et~al.}(2009){Lah}, {Pracy}, {Chengalur}, {Briggs}, {Colless},
  {de Propris}, {Ferris}, {Schmidt}, \& {Tucker}}]{lah2009}
{Lah}, P., {Pracy}, M.~B., {Chengalur}, J.~N., {et~al.} 2009, \mnras, 399, 1447

\bibitem[{{Lane}(2000)}]{lane2000}
{Lane}, W.~M. 2000, PhD thesis, University of Groningen

\bibitem[{{Lanzetta} {et~al.}(1987){Lanzetta}, {Turnshek}, \&
  {Wolfe}}]{lanzetta1987}
{Lanzetta}, K.~M., {Turnshek}, D.~A., \& {Wolfe}, A.~M. 1987, \apj, 322, 739

\bibitem[{{Liang} \& {Chen}(2014)}]{liang2014}
{Liang}, C.~J., \& {Chen}, H.-W. 2014, \mnras, 445, 2061

\bibitem[{{Liszt}(2001)}]{liszt2001}
{Liszt}, H. 2001, \aap, 371, 698

\bibitem[{{Maccagni} {et~al.}(2017){Maccagni}, {Morganti}, {Oosterloo},
  {Ger{\'e}b}, \& {Maddox}}]{maccagni2017}
{Maccagni}, F.~M., {Morganti}, R., {Oosterloo}, T.~A., {Ger{\'e}b}, K., \&
  {Maddox}, N. 2017, \aap, 604, A43

\bibitem[{{Madau} \& {Dickinson}(2014)}]{madau2014}
{Madau}, P., \& {Dickinson}, M. 2014, \araa, 52, 415

\bibitem[{{Masui} {et~al.}(2013){Masui}, {Switzer}, {Banavar}, {Bandura},
  {Blake}, {Calin}, {Chang}, {Chen}, {Li}, {Liao}, {Natarajan}, {Pen},
  {Peterson}, {Shaw}, \& {Voytek}}]{masui2013}
{Masui}, K.~W., {Switzer}, E.~R., {Banavar}, N., {et~al.} 2013, \apjl, 763, L20

\bibitem[{{McKee} \& {Ostriker}(1977)}]{mckee1977}
{McKee}, C.~F., \& {Ostriker}, J.~P. 1977, \apj, 218, 148

\bibitem[{{Mebold} {et~al.}(1997){Mebold}, {D{\"u}sterberg}, {Dickey},
  {Staveley-Smith}, \& {Kalberla}}]{mebold1997}
{Mebold}, U., {D{\"u}sterberg}, C., {Dickey}, J.~M., {Staveley-Smith}, L., \&
  {Kalberla}, P. 1997, \apjl, 490, L65

\bibitem[{{Meiring} {et~al.}(2011){Meiring}, {Tripp}, {Prochaska}, {Tumlinson},
  {Werk}, {Jenkins}, {Thom}, {O'Meara}, \& {Sembach}}]{meiring2011}
{Meiring}, J.~D., {Tripp}, T.~M., {Prochaska}, J.~X., {et~al.} 2011, \apj, 732,
  35

\bibitem[{{Mihos} {et~al.}(2012){Mihos}, {Keating}, {Holley-Bockelmann},
  {Pisano}, \& {Kassim}}]{mihos2012}
{Mihos}, J.~C., {Keating}, K.~M., {Holley-Bockelmann}, K., {Pisano}, D.~J., \&
  {Kassim}, N.~E. 2012, \apj, 761, 186

\bibitem[{{Morganti} \& {Oosterloo}(2018)}]{morganti2018}
{Morganti}, R., \& {Oosterloo}, T. 2018, \aapr, 26, 4

\bibitem[{{Muller} \& {Oort}(1951)}]{muller1951}
{Muller}, C.~A., \& {Oort}, J.~H. 1951, \nat, 168, 357

\bibitem[{{Neeleman} {et~al.}(2016){Neeleman}, {Prochaska}, {Ribaudo},
  {Lehner}, {Howk}, {Rafelski}, \& {Kanekar}}]{neeleman2016}
{Neeleman}, M., {Prochaska}, J.~X., {Ribaudo}, J., {et~al.} 2016, \apj, 818,
  113

\bibitem[{{Neeleman} {et~al.}(2015){Neeleman}, {Prochaska}, \&
  {Wolfe}}]{neeleman2015}
{Neeleman}, M., {Prochaska}, J.~X., \& {Wolfe}, A.~M. 2015, \apj, 800, 7

\bibitem[{{Nestor} {et~al.}(2005){Nestor}, {Turnshek}, \& {Rao}}]{nestor2005}
{Nestor}, D.~B., {Turnshek}, D.~A., \& {Rao}, S.~M. 2005, \apj, 628, 637

\bibitem[{{Nielsen} {et~al.}(2013){Nielsen}, {Churchill}, {Kacprzak}, \&
  {Murphy}}]{nielsen2013}
{Nielsen}, N.~M., {Churchill}, C.~W., {Kacprzak}, G.~G., \& {Murphy}, M.~T.
  2013, \apj, 776, 114

\bibitem[{{Noterdaeme} {et~al.}(2008){Noterdaeme}, {Ledoux}, {Petitjean}, \&
  {Srianand}}]{noterdaeme2008a}
{Noterdaeme}, P., {Ledoux}, C., {Petitjean}, P., \& {Srianand}, R. 2008, \aap,
  481, 327

\bibitem[{{Noterdaeme} {et~al.}(2009){Noterdaeme}, {Petitjean}, {Ledoux}, \&
  {Srianand}}]{noterdaeme2009b}
{Noterdaeme}, P., {Petitjean}, P., {Ledoux}, C., \& {Srianand}, R. 2009, \aap,
  505, 1087

\bibitem[{{Noterdaeme} {et~al.}(2012){Noterdaeme}, {Petitjean}, {Carithers},
  {P{\^a}ris}, {Font-Ribera}, {Bailey}, {Aubourg}, {Bizyaev}, {Ebelke},
  {Finley}, {Ge}, {Malanushenko}, {Malanushenko}, {Miralda-Escud{\'e}},
  {Myers}, {Oravetz}, {Pan}, {Pieri}, {Ross}, {Schneider}, {Simmons}, \&
  {York}}]{noterdaeme2012b}
{Noterdaeme}, P., {Petitjean}, P., {Carithers}, W.~C., {et~al.} 2012, \aap,
  547, L1

\bibitem[{{Oosterloo} {et~al.}(2007){Oosterloo}, {Fraternali}, \&
  {Sancisi}}]{oosterloo2007}
{Oosterloo}, T., {Fraternali}, F., \& {Sancisi}, R. 2007, \aj, 134, 1019

\bibitem[{{Patra} {et~al.}(2013){Patra}, {Chengalur}, \& {Begum}}]{patra2013}
{Patra}, N.~N., {Chengalur}, J.~N., \& {Begum}, A. 2013, \mnras, 429, 1596

\bibitem[{{Pawsey}(1951)}]{pawsey1951}
{Pawsey}, J.~L. 1951, \nat, 168, 358

\bibitem[{{P{\'e}roux} {et~al.}(2012){P{\'e}roux}, {Bouch{\'e}}, {Kulkarni},
  {York}, \& {Vladilo}}]{peroux2012}
{P{\'e}roux}, C., {Bouch{\'e}}, N., {Kulkarni}, V.~P., {York}, D.~G., \&
  {Vladilo}, G. 2012, \mnras, 419, 3060

\bibitem[{{P{\'e}roux} {et~al.}(2005){P{\'e}roux}, {Dessauges-Zavadsky},
  {D'Odorico}, {Sun Kim}, \& {McMahon}}]{peroux2005}
{P{\'e}roux}, C., {Dessauges-Zavadsky}, M., {D'Odorico}, S., {Sun Kim}, T., \&
  {McMahon}, R.~G. 2005, \mnras, 363, 479

\bibitem[{{P{\'e}roux} {et~al.}(2019){P{\'e}roux}, {Zwaan}, {Klitsch},
  {Augustin}, {Hamanowicz}, {Rahmani}, {Pettini}, {Kulkarni}, {Straka},
  {Biggs}, {York}, \& {Milliard}}]{peroux2019}
{P{\'e}roux}, C., {Zwaan}, M.~A., {Klitsch}, A., {et~al.} 2019, \mnras, 485,
  1595

\bibitem[{{Pontzen} {et~al.}(2008){Pontzen}, {Governato}, {Pettini}, {Booth},
  {Stinson}, {Wadsley}, {Brooks}, {Quinn}, \& {Haehnelt}}]{pontzen2008}
{Pontzen}, A., {Governato}, F., {Pettini}, M., {et~al.} 2008, \mnras, 390, 1349

\bibitem[{{Prochaska} {et~al.}(2005){Prochaska}, {Herbert-Fort}, \&
  {Wolfe}}]{prochaska2005}
{Prochaska}, J.~X., {Herbert-Fort}, S., \& {Wolfe}, A.~M. 2005, \apj, 635, 123

\bibitem[{{Prochaska} {et~al.}(2011){Prochaska}, {Weiner}, {Chen}, {Mulchaey},
  \& {Cooksey}}]{prochaska2011}
{Prochaska}, J.~X., {Weiner}, B., {Chen}, H.-W., {Mulchaey}, J., \& {Cooksey},
  K. 2011, \apj, 740, 91

\bibitem[{{Prochter} {et~al.}(2006){Prochter}, {Prochaska}, \&
  {Burles}}]{prochter2006}
{Prochter}, G.~E., {Prochaska}, J.~X., \& {Burles}, S.~M. 2006, \apj, 639, 766

\bibitem[{{Quider} {et~al.}(2011){Quider}, {Nestor}, {Turnshek}, {Rao},
  {Monier}, {Weyant}, \& {Busche}}]{quider2011}
{Quider}, A.~M., {Nestor}, D.~B., {Turnshek}, D.~A., {et~al.} 2011, \aj, 141,
  137

\bibitem[{{Rahmani} {et~al.}(2012){Rahmani}, {Srianand}, {Gupta}, {Petitjean},
  {Noterdaeme}, \& {V{\'a}squez}}]{rahmani2012}
{Rahmani}, H., {Srianand}, R., {Gupta}, N., {et~al.} 2012, \mnras, 425, 556

\bibitem[{{Rahmani} {et~al.}(2016){Rahmani}, {P{\'e}roux}, {Turnshek}, {Rao},
  {Quiret}, {Hamilton}, {Kulkarni}, {Monier}, \& {Zafar}}]{rahmani2016}
{Rahmani}, H., {P{\'e}roux}, C., {Turnshek}, D.~A., {et~al.} 2016, \mnras, 463,
  980

\bibitem[{{Rao} {et~al.}(2011){Rao}, {Belfort-Mihalyi}, {Turnshek}, {Monier},
  {Nestor}, \& {Quider}}]{rao2011}
{Rao}, S.~M., {Belfort-Mihalyi}, M., {Turnshek}, D.~A., {et~al.} 2011, \mnras,
  416, 1215

\bibitem[{{Rao} {et~al.}(2006){Rao}, {Turnshek}, \& {Nestor}}]{rao2006}
{Rao}, S.~M., {Turnshek}, D.~A., \& {Nestor}, D.~B. 2006, \apj, 636, 610

\bibitem[{{Rao} {et~al.}(2017){Rao}, {Turnshek}, {Sardane}, \&
  {Monier}}]{rao2017}
{Rao}, S.~M., {Turnshek}, D.~A., {Sardane}, G.~M., \& {Monier}, E.~M. 2017,
  \mnras, 471, 3428

\bibitem[{{Reeves} {et~al.}(2016){Reeves}, {Sadler}, {Allison}, {Koribalski},
  {Curran}, {Pracy}, {Phillips}, {Bignall}, \& {Reynolds}}]{reeves2016}
{Reeves}, S.~N., {Sadler}, E.~M., {Allison}, J.~R., {et~al.} 2016, \mnras, 457,
  2613

\bibitem[{{Rhee} {et~al.}(2018){Rhee}, {Lah}, {Briggs}, {Chengalur}, {Colless},
  {Willner}, {Ashby}, \& {Le F{\`e}vre}}]{rhee2018}
{Rhee}, J., {Lah}, P., {Briggs}, F.~H., {et~al.} 2018, \mnras, 473, 1879

\bibitem[{{Rhee} {et~al.}(2013){Rhee}, {Zwaan}, {Briggs}, {Chengalur}, {Lah},
  {Oosterloo}, \& {van der Hulst}}]{rhee2013}
{Rhee}, J., {Zwaan}, M.~A., {Briggs}, F.~H., {et~al.} 2013, \mnras, 435, 2693

\bibitem[{{Roberts}(1970)}]{roberts1970}
{Roberts}, M.~S. 1970, \apjl, 161, L9

\bibitem[{{Rohlfs} \& {Wilson}(2000)}]{rohlfs2000}
{Rohlfs}, K., \& {Wilson}, T.~L. 2000, {Tools of radio astronomy}

\bibitem[{{Rosenberg} \& {Schneider}(2002)}]{rosenberg2002}
{Rosenberg}, J.~L., \& {Schneider}, S.~E. 2002, \apj, 567, 247

\bibitem[{{Roy} {et~al.}(2006){Roy}, {Chengalur}, \& {Srianand}}]{roy2006}
{Roy}, N., {Chengalur}, J.~N., \& {Srianand}, R. 2006, \mnras, 365, L1

\bibitem[{{Roy} {et~al.}(2013){Roy}, {Kanekar}, \& {Chengalur}}]{roy2013}
{Roy}, N., {Kanekar}, N., \& {Chengalur}, J.~N. 2013, \mnras, 436, 2366

\bibitem[{{Rubin} {et~al.}(1985){Rubin}, {Burstein}, {Ford}, \&
  {Thonnard}}]{rubin1985}
{Rubin}, V.~C., {Burstein}, D., {Ford}, Jr., W.~K., \& {Thonnard}, N. 1985,
  \apj, 289, 81

\bibitem[{{S{\'a}nchez} {et~al.}(2012){S{\'a}nchez}, {Kennicutt}, {Gil de Paz},
  {van de Ven}, {V{\'{\i}}lchez}, {Wisotzki}, {Walcher}, {Mast}, {Aguerri},
  {Albiol-P{\'e}rez}, {Alonso-Herrero}, {Alves}, {Bakos}, {Bart{\'a}kov{\'a}},
  {Bland-Hawthorn}, {Boselli}, {Bomans}, {Castillo-Morales}, {Cortijo-Ferrero},
  {de Lorenzo-C{\'a}ceres}, {Del Olmo}, {Dettmar}, {D{\'{\i}}az}, {Ellis},
  {Falc{\'o}n-Barroso}, {Flores}, {Gallazzi}, {Garc{\'{\i}}a-Lorenzo},
  {Gonz{\'a}lez Delgado}, {Gruel}, {Haines}, {Hao}, {Husemann},
  {Igl{\'e}sias-P{\'a}ramo}, {Jahnke}, {Johnson}, {Jungwiert}, {Kalinova},
  {Kehrig}, {Kupko}, {L{\'o}pez-S{\'a}nchez}, {Lyubenova}, {Marino},
  {M{\'a}rmol-Queralt{\'o}}, {M{\'a}rquez}, {Masegosa}, {Meidt},
  {Mendez-Abreu}, {Monreal-Ibero}, {Montijo}, {Mour{\~a}o}, {Palacios-Navarro},
  {Papaderos}, {Pasquali}, {Peletier}, {P{\'e}rez}, {P{\'e}rez}, {Quirrenbach},
  {Rela{\~n}o}, {Rosales-Ortega}, {Roth}, {Ruiz-Lara},
  {S{\'a}nchez-Bl{\'a}zquez}, {Sengupta}, {Singh}, {Stanishev}, {Trager},
  {Vazdekis}, {Viironen}, {Wild}, {Zibetti}, \& {Ziegler}}]{sanchez2012}
{S{\'a}nchez}, S.~F., {Kennicutt}, R.~C., {Gil de Paz}, A., {et~al.} 2012,
  \aap, 538, A8

\bibitem[{{Sancisi} {et~al.}(2008){Sancisi}, {Fraternali}, {Oosterloo}, \& {van
  der Hulst}}]{sancisi2008}
{Sancisi}, R., {Fraternali}, F., {Oosterloo}, T., \& {van der Hulst}, T. 2008,
  \aapr, 15, 189

\bibitem[{{Sargent} {et~al.}(1988){Sargent}, {Steidel}, \&
  {Boksenberg}}]{sargent1988}
{Sargent}, W.~L.~W., {Steidel}, C.~C., \& {Boksenberg}, A. 1988, \apj, 334, 22

\bibitem[{{Shuter} \& {Gower}(1969)}]{shuter1969}
{Shuter}, W.~L.~H., \& {Gower}, J.~F.~R. 1969, \nat, 223, 1046

\bibitem[{{Srianand} {et~al.}(2012){Srianand}, {Gupta}, {Petitjean},
  {Noterdaeme}, {Ledoux}, {Salter}, \& {Saikia}}]{srianand2012}
{Srianand}, R., {Gupta}, N., {Petitjean}, P., {et~al.} 2012, \mnras, 421, 651

\bibitem[{{Srianand} {et~al.}(2013){Srianand}, {Gupta}, {Rahmani}, {Momjian},
  {Petitjean}, \& {Noterdaeme}}]{srianand2013}
{Srianand}, R., {Gupta}, N., {Rahmani}, H., {et~al.} 2013, \mnras, 428, 2198

\bibitem[{{Steidel}(1995)}]{steidel1995}
{Steidel}, C.~C. 1995, in QSO Absorption Lines, ed. G.~{Meylan}, 139

\bibitem[{{Steidel} \& {Sargent}(1992)}]{steidel1992}
{Steidel}, C.~C., \& {Sargent}, W.~L.~W. 1992, \apjs, 80, 1

\bibitem[{{Stocke} {et~al.}(2013){Stocke}, {Keeney}, {Danforth}, {Shull},
  {Froning}, {Green}, {Penton}, \& {Savage}}]{stocke2013}
{Stocke}, J.~T., {Keeney}, B.~A., {Danforth}, C.~W., {et~al.} 2013, \apj, 763,
  148

\bibitem[{{Tamburro} {et~al.}(2009){Tamburro}, {Rix}, {Leroy}, {Mac Low},
  {Walter}, {Kennicutt}, {Brinks}, \& {de Blok}}]{tamburro2009}
{Tamburro}, D., {Rix}, H.-W., {Leroy}, A.~K., {et~al.} 2009, \aj, 137, 4424

\bibitem[{{Tumlinson} {et~al.}(2017){Tumlinson}, {Peeples}, \&
  {Werk}}]{tumlinson2017}
{Tumlinson}, J., {Peeples}, M.~S., \& {Werk}, J.~K. 2017, \araa, 55, 389

\bibitem[{{Tumlinson} {et~al.}(2011){Tumlinson}, {Thom}, {Werk}, {Prochaska},
  {Tripp}, {Weinberg}, {Peeples}, {O'Meara}, {Oppenheimer}, {Meiring}, {Katz},
  {Dav{\'e}}, {Ford}, \& {Sembach}}]{tumlinson2011}
{Tumlinson}, J., {Thom}, C., {Werk}, J.~K., {et~al.} 2011, Science, 334, 948

\bibitem[{{Tumlinson} {et~al.}(2013){Tumlinson}, {Thom}, {Werk}, {Prochaska},
  {Tripp}, {Katz}, {Dav{\'e}}, {Oppenheimer}, {Meiring}, {Ford}, {O'Meara},
  {Peeples}, {Sembach}, \& {Weinberg}}]{tumlinson2013}
---. 2013, \apj, 777, 59

\bibitem[{{Turnshek} {et~al.}(2015){Turnshek}, {Monier}, {Rao}, {Hamilton},
  {Sardane}, \& {Held}}]{turnshek2015}
{Turnshek}, D.~A., {Monier}, E.~M., {Rao}, S.~M., {et~al.} 2015, \mnras, 449,
  1536

\bibitem[{{van Albada} {et~al.}(1985){van Albada}, {Bahcall}, {Begeman}, \&
  {Sancisi}}]{vanalbada1985}
{van Albada}, T.~S., {Bahcall}, J.~N., {Begeman}, K., \& {Sancisi}, R. 1985,
  \apj, 295, 305

\bibitem[{{van de Hulst}(1945)}]{vandehulst1945}
{van de Hulst}, H.~C. 1945, Nederl. Tij. Natuurkunde, 11, 201

\bibitem[{{van der Hulst} {et~al.}(2001){van der Hulst}, {van Albada}, \&
  {Sancisi}}]{vanderhulst2001}
{van der Hulst}, J.~M., {van Albada}, T.~S., \& {Sancisi}, R. 2001, in
  Astronomical Society of the Pacific Conference Series, Vol. 240, Gas and
  Galaxy Evolution, ed. J.~E. {Hibbard}, M.~{Rupen}, \& J.~H. {van Gorkom}, 451

\bibitem[{{Verheijen} {et~al.}(2007){Verheijen}, {van Gorkom}, {Szomoru},
  {Dwarakanath}, {Poggianti}, \& {Schiminovich}}]{verheijen2007}
{Verheijen}, M., {van Gorkom}, J.~H., {Szomoru}, A., {et~al.} 2007, \apjl, 668,
  L9

\bibitem[{{Wakker} {et~al.}(2011){Wakker}, {Lockman}, \& {Brown}}]{wakker2011}
{Wakker}, B.~P., {Lockman}, F.~J., \& {Brown}, J.~M. 2011, \apj, 728, 159

\bibitem[{{Walter} {et~al.}(2008){Walter}, {Brinks}, {de Blok}, {Bigiel},
  {Kennicutt}, {Thornley}, \& {Leroy}}]{walter2008}
{Walter}, F., {Brinks}, E., {de Blok}, W.~J.~G., {et~al.} 2008, \aj, 136, 2563

\bibitem[{{Werk} {et~al.}(2014){Werk}, {Prochaska}, {Tumlinson}, {Peeples},
  {Tripp}, {Fox}, {Lehner}, {Thom}, {O'Meara}, {Ford}, {Bordoloi}, {Katz},
  {Tejos}, {Oppenheimer}, {Dav{\'e}}, \& {Weinberg}}]{werk2014}
{Werk}, J.~K., {Prochaska}, J.~X., {Tumlinson}, J., {et~al.} 2014, \apj, 792, 8

\bibitem[{{Wolfe} {et~al.}(1981){Wolfe}, {Briggs}, \& {Jauncey}}]{wolfe1981}
{Wolfe}, A.~M., {Briggs}, F.~H., \& {Jauncey}, D.~L. 1981, \apj, 248, 460

\bibitem[{{Wolfe} {et~al.}(1985){Wolfe}, {Briggs}, {Turnshek}, {Davis},
  {Smith}, \& {Cohen}}]{wolfe1985}
{Wolfe}, A.~M., {Briggs}, F.~H., {Turnshek}, D.~A., {et~al.} 1985, \apjl, 294,
  L67

\bibitem[{{Wolfe} {et~al.}(1976){Wolfe}, {Brown}, \& {Roberts}}]{wolfe1976}
{Wolfe}, A.~M., {Brown}, R.~L., \& {Roberts}, M.~S. 1976, \prl, 37, 179

\bibitem[{{Wolfe} \& {Davis}(1979)}]{wolfe1979}
{Wolfe}, A.~M., \& {Davis}, M.~M. 1979, \aj, 84, 699

\bibitem[{{Wolfe} {et~al.}(2005){Wolfe}, {Gawiser}, \& {Prochaska}}]{wolfe2005}
{Wolfe}, A.~M., {Gawiser}, E., \& {Prochaska}, J.~X. 2005, \araa, 43, 861

\bibitem[{{Wolfire} {et~al.}(1995){Wolfire}, {Hollenbach}, {McKee}, {Tielens},
  \& {Bakes}}]{wolfire1995}
{Wolfire}, M.~G., {Hollenbach}, D., {McKee}, C.~F., {Tielens}, A.~G.~G.~M., \&
  {Bakes}, E.~L.~O. 1995, \apj, 443, 152

\bibitem[{{Wolfire} {et~al.}(2003){Wolfire}, {McKee}, {Hollenbach}, \&
  {Tielens}}]{wolfire2003}
{Wolfire}, M.~G., {McKee}, C.~F., {Hollenbach}, D., \& {Tielens}, A.~G.~G.~M.
  2003, \apj, 587, 278

\bibitem[{{York} {et~al.}(2000){York}, {Adelman}, {Anderson}, {Anderson},
  {Annis}, {Bahcall}, {Bakken}, {Barkhouser}, {Bastian}, {Berman}, {Boroski},
  {Bracker}, {Briegel}, {Briggs}, {Brinkmann}, {Brunner}, {Burles}, {Carey},
  {Carr}, {Castander}, {Chen}, {Colestock}, {Connolly}, {Crocker}, {Csabai},
  {Czarapata}, {Davis}, {Doi}, {Dombeck}, {Eisenstein}, {Ellman}, {Elms},
  {Evans}, {Fan}, {Federwitz}, {Fiscelli}, {Friedman}, {Frieman}, {Fukugita},
  {Gillespie}, {Gunn}, {Gurbani}, {de Haas}, {Haldeman}, {Harris}, {Hayes},
  {Heckman}, {Hennessy}, {Hindsley}, {Holm}, {Holmgren}, {Huang}, {Hull},
  {Husby}, {Ichikawa}, {Ichikawa}, {Ivezi{\'c}}, {Kent}, {Kim}, {Kinney},
  {Klaene}, {Kleinman}, {Kleinman}, {Knapp}, {Korienek}, {Kron}, {Kunszt},
  {Lamb}, {Lee}, {Leger}, {Limmongkol}, {Lindenmeyer}, {Long}, {Loomis},
  {Loveday}, {Lucinio}, {Lupton}, {MacKinnon}, {Mannery}, {Mantsch}, {Margon},
  {McGehee}, {McKay}, {Meiksin}, {Merelli}, {Monet}, {Munn}, {Narayanan},
  {Nash}, {Neilsen}, {Neswold}, {Newberg}, {Nichol}, {Nicinski}, {Nonino},
  {Okada}, {Okamura}, {Ostriker}, {Owen}, {Pauls}, {Peoples}, {Peterson},
  {Petravick}, {Pier}, {Pope}, {Pordes}, {Prosapio}, {Rechenmacher}, {Quinn},
  {Richards}, {Richmond}, {Rivetta}, {Rockosi}, {Ruthmansdorfer}, {Sandford},
  {Schlegel}, {Schneider}, {Sekiguchi}, {Sergey}, {Shimasaku}, {Siegmund},
  {Smee}, {Smith}, {Snedden}, {Stone}, {Stoughton}, {Strauss}, {Stubbs},
  {SubbaRao}, {Szalay}, {Szapudi}, {Szokoly}, {Thakar}, {Tremonti}, {Tucker},
  {Uomoto}, {Vanden Berk}, {Vogeley}, {Waddell}, {Wang}, {Watanabe},
  {Weinberg}, {Yanny}, {Yasuda}, \& {SDSS Collaboration}}]{york2000}
{York}, D.~G., {Adelman}, J., {Anderson}, Jr., J.~E., {et~al.} 2000, \aj, 120,
  1579

\bibitem[{{Zhu} \& {M{\'e}nard}(2013)}]{zhu2013}
{Zhu}, G., \& {M{\'e}nard}, B. 2013, \apj, 770, 130

\bibitem[{{Zwaan} {et~al.}(2015){Zwaan}, {Liske}, {P{\'e}roux}, {Murphy},
  {Bouch{\'e}}, {Curran}, \& {Biggs}}]{zwaan2015}
{Zwaan}, M.~A., {Liske}, J., {P{\'e}roux}, C., {et~al.} 2015, \mnras, 453, 1268

\bibitem[{{Zwaan} {et~al.}(2005{\natexlab{a}}){Zwaan}, {Meyer},
  {Staveley-Smith}, \& {Webster}}]{zwaan2005a}
{Zwaan}, M.~A., {Meyer}, M.~J., {Staveley-Smith}, L., \& {Webster}, R.~L.
  2005{\natexlab{a}}, \mnras, 359, L30

\bibitem[{{Zwaan} {et~al.}(2005{\natexlab{b}}){Zwaan}, {van der Hulst},
  {Briggs}, {Verheijen}, \& {Ryan-Weber}}]{zwaan2005}
{Zwaan}, M.~A., {van der Hulst}, J.~M., {Briggs}, F.~H., {Verheijen}, M.~A.~W.,
  \& {Ryan-Weber}, E.~V. 2005{\natexlab{b}}, \mnras, 364, 1467

\bibitem[{{Zwaan} {et~al.}(2001){Zwaan}, {van Dokkum}, \&
  {Verheijen}}]{zwaan2001}
{Zwaan}, M.~A., {van Dokkum}, P.~G., \& {Verheijen}, M.~A.~W. 2001, Science,
  293, 1800

\end{thebibliography}

\end{document}